\documentclass[reprint,prb,aps,twocolumn,superscriptaddress]{revtex4-1}

\usepackage{bm}
\usepackage{ulem}
\usepackage{graphicx}
\usepackage{times}
\usepackage{soul}
\usepackage{color}
\usepackage{braket}
\usepackage{physics}
\usepackage{natbib}
\usepackage{floatrow}
\usepackage{physics}
\usepackage{comment}

\newcommand{\ba}{\begin{align}}
\newcommand{\ea}{\end{align}}
\newcommand{\B}{\mathbf}

\usepackage[colorlinks,
            citecolor = blue,
						linkcolor = blue,
						urlcolor  = blue,
						anchorcolor = blue,
						breaklinks = true
						]{hyperref}

\def \ETH{Institute for Quantum Electronics, ETH Z\"urich, CH-8093 Z\"urich, Switzerland}
\def \MPI{Max Planck Institute of Quantum Optics, 85748 Garching, Germany}
\def \MCQST{Munich Center for Quantum Science and Technology, Schellingstrasse 4, 80799 M\"unich, Germany}

\begin{document}

\title{Interacting polaron-polaritons}

\author{Li Bing Tan}\affiliation{\ETH}
\author{Ovidiu Cotlet}\affiliation{\ETH}
\author{Andrea Bergschneider}\affiliation{\ETH}
\author{Richard Schmidt}\affiliation{\MPI}\affiliation{\MCQST}
\author{Patrick Back}\affiliation{\ETH}
\author{Yuya Shimazaki}\affiliation{\ETH}
\author{Martin Kroner}\affiliation{\ETH}
\author{Atac Imamoglu}\email{imamoglu@phys.ethz.ch}\affiliation{\ETH}

\begin{abstract}
    Two dimensional semiconductors provide an ideal platform for exploration of linear exciton and polariton physics, primarily due to large exciton binding energy and strong light-matter coupling. These features, however, generically imply reduced exciton-exciton interactions, hindering the realisation of active optical devices such as lasers or parametric oscillators. Here, we show that electrical injection of itinerant electrons into monolayer molybdenum diselenide allows us to overcome this limitation: dynamical screening of exciton-polaritons by electrons leads to the formation of new quasi-particles termed polaron-polaritons that exhibit unexpectedly strong interactions as well as optical amplification by Bose-enhanced polaron-electron scattering. To measure the nonlinear optical response, we carry out time-resolved pump-probe measurements and observe polaron-polariton interaction enhancement by a factor of 50 ($0.5 \mu$eV $\mu$m$^2$) as compared to exciton-polaritons. Concurrently, we measure a spectrally integrated transmission gain of the probe field of $\gtrsim 2$ stemming from stimulated scattering of polaron-polaritons. We show theoretically that the non-equilibrium nature of optically excited quasiparticles favours a previously unexplored interaction mechanism stemming from a phase-space filling in the screening cloud, which provides an accurate explanation of the strong repulsive interactions observed experimentally. Our findings show that itinerant electron-exciton interactions provide an invaluable tool for electronic manipulation of optical properties, demonstrate a new mechanism for dramatically enhancing polariton-polariton interactions, and pave the way for realisation of nonequilibrium polariton condensates.
\end{abstract}

\maketitle

\section{Introduction}
Monolayer transition metal dichalcogenides (TMDs) combine strong light-matter coupling and a large number of degrees of freedom (DOF) with which to manipulate photons. Due to the strong Coulomb interaction, excitons constitute the elementary optical excitation and dominate optical spectra in TMDs. Many novel features of linear optical properties of TMD excitons have been demonstrated, including the optical control of the valley DOF using the helicity of the excitation light~\cite{Mak2012, Zeng2012} or external magnetic fields~\cite{Srivastava2015,Aivazian2015,Smolenski2016}, electrical control of the optical spectrum through injection of itinerant electrons or holes~\cite{Xu16-NatRevMat}, and the realisation of atomically-thin mirrors~\cite{Back18-PRL,Scuri18-PRL}. However, nonlinear optical properties, crucial for the realisation of novel photonic devices ranging from parametric oscillators to exciton-polariton lasers, have been largely missing in the reported studies. This is at a first glance not surprising since nonlinearities typically scale as the exciton Bohr radius and hence are suppressed in TMD monolayers where excitons are strongly bound~\cite{Heinz14-NatPhys,Xu16-NatRevMat,Urbaszek18-RMP}, as compared to quasi-2D materials such as gallium arsenide (GaAs) quantum wells. Consequently, an outstanding challenge for the field is to find ways to enhance the exciton-exciton interaction strength, or more precisely its ratio to the exciton radiative decay rate. Theoretical proposals to date include the exploitation of large interactions between Rydberg excitons with three-level driving schemes~\cite{Walter2018} and to reduce the exciton radiative decay rate by placing  a monolayer in a high quality-factor (Q) cavity~\cite{zeytinoglu18} or, more elegantly, by placing a monolayer at $\lambda/2$ distance away from a mirror~\cite{Wild2018}. However, experiments have so far only confirmed that exciton-exciton interactions in neutral TMD monolayers are more than an order of magnitude weaker than those in GaAs~\cite{Barachati2018,Scuri18-PRL}.

In this work, we demonstrate that the presence of itinerant electrons dramatically enhances nonlinear optical properties of TMD monolayers. In an earlier work, Sidler et al.~\cite{MSidler2017} investigated the linear optics of doped monolayer MoSe$_2$  and identified exciton-polarons as the relevant quasiparticles. In TMD monolayers, the exciton has an ultra large binding energy that dominates over the electron plasma frequency and Fermi energy. In consequence, the exciton can be considered as a robust quantum impurity interacting with a fermionic bath. The itinerant electrons dynamically screen the exciton to form new quasiparticle branches -- the attractive and repulsive polaron -- each with a renormalised mass and energy~\cite{MSidler2017,Macdonald17}. A simple description of the polaron as a superposition of a bare exciton and an exciton dressed with a single electron-hole pair~\cite{chevy2006universal} was sufficient to accurately predict the resonances observed in linear spectroscopy.

Here, we use nonlinear spectroscopy to investigate the residual quasiparticle interactions in the strong cavity-coupling regime~\cite{weisbuch92-PRL,deng2010exciton,Carusotto2013,Menon2015,Tartakovskii2015} where elementary optical excitations are polaron-polaritons~\cite{MSidler2017}. We find their effective polariton-polariton interaction strength to exceed that of their undressed counterparts by up to a factor of $\sim 50$ , and we demonstrate an unexpected but unequivocal amplification of polaron-polaritons accompanying the interaction induced blueshifts, with gain factors $\gtrsim 2$. 

The enhanced nonlinearities can be attributed to residual interactions between the polaron quasiparticles. Using a wavefunction technique, we show how the measured repulsive interaction shift in the presence of itinerant electrons can be understood in terms of a phase-space filling effect (PSF). Here, the strong correlations between pump-generated polaritons and electrons, that are associated with the formation of a polaron dressing cloud, lead to an effective depletion of the electronic medium. Thus, additional polaritons created by the probe cannot be screened with maximal efficiency, resulting in an increase of their energy. Already at the lowest order, our approach leads to a remarkable agreement between theory and experiment: we attribute this in part to the zero-dimensional nature of the cavity mode which ensures that the lowest energy optically excited state of the coupled system -- the lower-polaron-polariton (polaron-LP) -- is effectively gapped from the higher-lying continuum of polaron states, thereby suppressing higher-order interaction processes.

Residual interactions between polarons and electrons are also responsible for the observed polariton amplification~\cite{kavokin02-PRB,Baumberg03-PRL,Bloch05-PRB,Tartakovskii03-PRB,Qarry03-PRB}. A fraction of pump-generated polaritons contribute to the creation of a high-momentum polaron reservoir. These polarons then relax into the polaron-LP state~\cite{Bruun10-PRL,Schmidt2011-PRA} by generating additional electron-hole pair excitations in the electron system. In the presence of a seed population in the polaron-LP state, this scattering process is Bose enhanced~\cite{imamoglu96-PRA,baumberg00-PRL}, leading to the observed gain.

After describing the experimental setup and reviewing the linear optical properties in the presence of non-perturbative coupling between excitons and the cavity mode as well as excitons and electrons in Section~\ref{sec:two}, we present the experimental signatures unveiled in nonlinear spectroscopy, namely the enhancement of nonlinearities (Section~\ref{sec:three}) and polariton amplification in the presence of itinerant electrons as compared to the charge neutral regime (Section~\ref{sec:four}). We then present the theoretical model and calculations of the scattering rates due to residual interactions in Section~\ref{sec:five}. Finally, in Section~\ref{sec:six}, we discuss avenues for further work.

\section{Elementary optical excitations of an electron-doped monolayer \texorpdfstring{MoSe$_2$}{MoSe2} embedded in a microcavity}
\label{sec:two}

\begin{figure}
            \includegraphics[width=\textwidth]{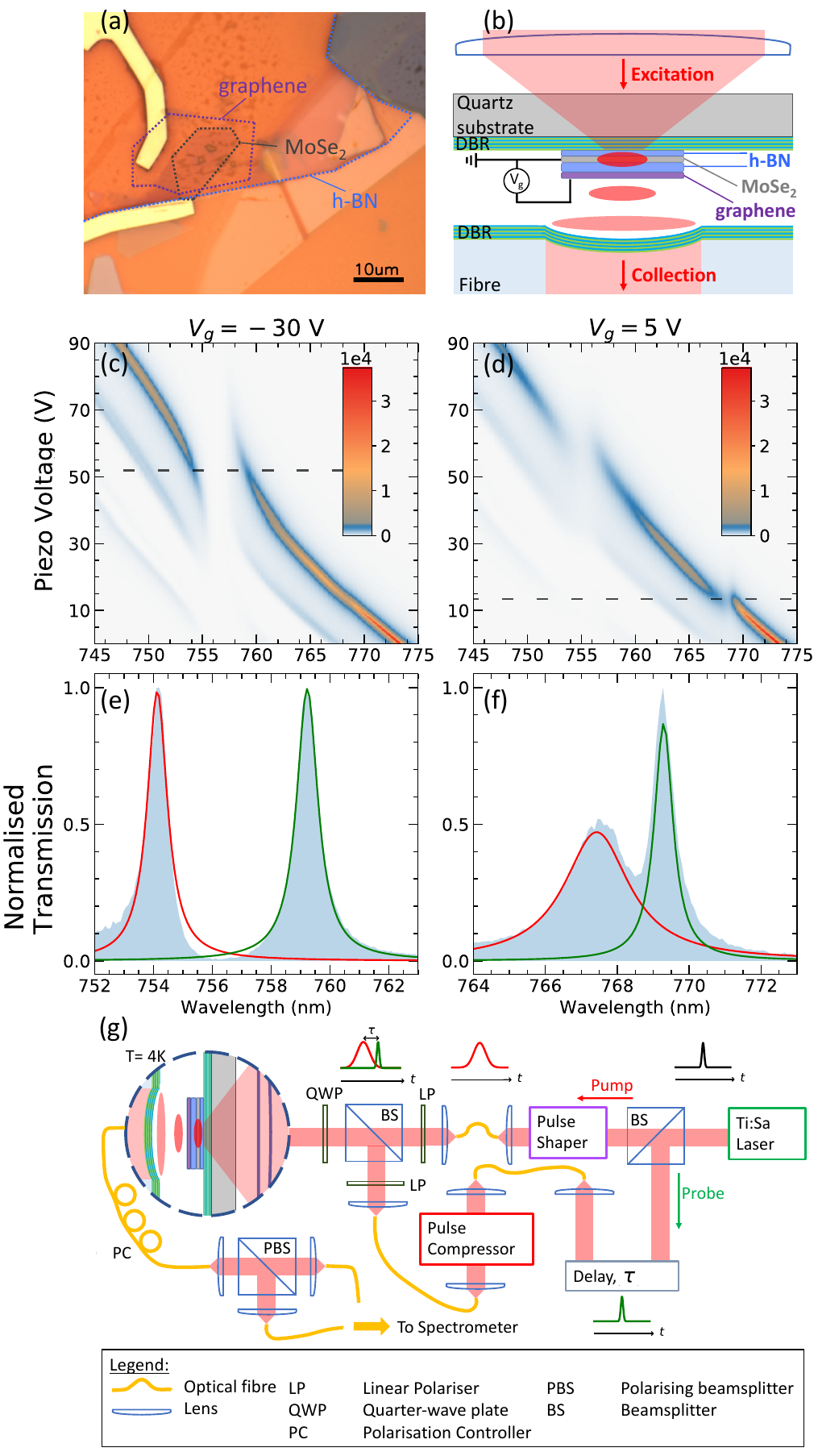}
    \caption{ Experimental setup and characterization of the sample. (a) Contrast-enhanced optical microscope image of the heterostructure. The h-BN, graphene and MoSe$_{2}$ layers are indicated by dotted lines. (b) Schematic of the sample inside the cavity. The heterostructure sits on a flat dielectric mirror facing a fibre mirror which together form a cavity. The h-BN and graphene layer thicknesses are chosen such that the MoSe$_{2}$ lies in an antinode while the graphene lies close to a node of the cavity electric field. Applying a gate voltage between MoSe$_{2}$ and graphene provides electron density control. The cavity mode can be tuned in-situ by a piezo which controls the cavity length. (c), (d) White light transmission spectrum for gate voltages $V_{\mathrm{g}}= -30$~V and $V_{\mathrm{g}} =5$~V, respectively. Non-perturbative coupling between the cavity mode and optical resonances show up as anti-crossings in the transmission spectrum. (e), (f) Linecuts taken near the anti-crossing (dashed lines in (c) and (d)) reveal the upper polariton (red) and lower polariton (green) resonances. (g) A schematic of the pump-probe setup. From right to left: The output of a mode-locked Ti:Sapphire laser (76 MHz repetition rate) is split into two arms: pump and probe. The pump is spectrally filtered using a pulse shaper setup. The probe pulse is sent through an optical delay line to control the time delay $\tau$ with respect to the pump pulse, and a pulse compressor to compensate for linear dispersion in the optical fibres.
    }
\label{fig: figure1}
\end{figure}

Our experiments are based on a van der Waals heterostructure consisting of a MoSe$_{2}$ monolayer semiconductor embedded between 2 hexagonal boron nitride (hBN) flakes. A top graphene layer allows us to control the electron density (Fig~\ref{fig: figure1}a). We place the heterostructure in a tunable zero-dimensional open cavity~\cite{Besga15-PRAppl} consisting of a flat DBR-coated transparent substrate and a concave DBR-coated fibre facet (Fig~\ref{fig: figure1}b) ($Q \simeq 1800$). All experiments are carried out at liquid Helium temperature (see Appendix A).

Figure~\ref{fig: figure1}c and~\ref{fig: figure1}d show the white light transmission spectrum measured for vanishing electron density ($V_{\mathrm{g}} = -30$~V) and finite electron density ($V_{\mathrm{g}} = 5$~V), respectively, in the strong coupling regime as a function of the cavity length. In both cases, the measured spectra exhibit avoided crossings associated with the formation of half-light, half-matter quasiparticles termed polaritons~\cite{MSidler2017,Schneider2018}. In the charge neutral regime at $V_{\mathrm{g}} = -30$~V, we observe an exciton-polariton normal mode splitting of 14 meV. When the monolayer is electron-doped ($V_{\mathrm{g}} = 5$~V), dynamical screening of the excitons by electrons dramatically alters the nature of elementary optical excitations, leading to the appearance of attractive and repulsive exciton-polaron resonances. 

To analyse how the exciton-electron interaction modifies the ground state of the cavity-coupled system, we start from the Hamiltonian $H = H_{xe} + H_{cav}$ which can be written as
\begin{eqnarray}
\label{eqn:hamiltonian}
    H_{xe} = \sum_{\B k}\omega_\B k x^{\dagger}_{\B k} x_{\B k} &+& \sum_{\B k} \epsilon_\B k e^{\dagger}_{\B k} e_{\B k}\\
    \nonumber
    &+& \frac{v}{\cal A}\sum_{\B {k,k',q}}  x^{\dagger}_{\B {k+q}} x_{\B k} e^{\dagger}_{\B {k'-q}} e_{\B k'},\\
\label{eqn:hamiltonian2}
    H_{cav} = \sum_{\B k} \nu_\B k c^{\dagger}_{\B k} c_{\B k} &+& \sum_{\B k}  \Omega (c^{\dagger}_{\B k} x_{\B k} + h.c.).
\end{eqnarray}
Here, $x_{\B k}$, $e_{\B k}$ and $c_{\B k}$ are the annihilation operators of the exciton, electron and cavity photon of momentum $\B k$ respectively, with $\omega_\B k=|\B k|^2/(2 m_x)$, $\epsilon_\B k=|\B k|^2/(2 m_e)$ and $\nu_\B k=|\B k|^2/(2 m_c)+\Delta$ their energy dispersions. The detuning $\Delta$ between cavity photons and the excitons is controlled by changing the cavity length. For numerics we will use the electron mass $m_e=0.6m_0$, the exciton mass $m_x=2 m_e$ while the cavity mass $m_c \approx 10^{-5}m_0$.

Notice that in the absence of doping, the first term in Eq.~\ref{eqn:hamiltonian} describes excitons as elementary excitations in the MoSe$_2$ that exhibit weak mutual residual interactions  which we have chosen to neglect here. Including the second and third term takes into account the itinerant electrons and their effective interaction $v$ with excitons when the monolayer is capacitively doped, which we model here as a contact interaction and $\cal A$ is defined as the quantisation area. Eq.~\ref{eqn:hamiltonian2} describes the effect of the cavity mode: the first term takes into account the cavity energy and the second term which is proportional to $\Omega$ denotes the exciton-cavity coupling. 

The contact interaction $v$ between excitons and electrons is regularized by a UV cutoff $\Lambda$. Physically, $\Lambda$ can be related to the inverse Bohr radius of the exciton. However, assuming that the exciton Bohr radius is the smallest lengthscale in the problem, one may take $\Lambda \to \infty$ at the end of the calculation. Since any attractive interaction supports a bound state in 2D, the constant $v$ can be related to the binding energy of the exciton-electron bound state known as the trion: 
\begin{align}
    v^{-1}=-\frac{1}{\cal A}\sum_{|\B k|<\Lambda} \frac{1}{E_T + \omega_\B k + \epsilon_\B k}
\end{align}
where $E_T$ denotes the trion binding energy. 

It has been shown that the eigenstates of the interacting polariton-electron system can be accurately described using the Chevy Ansatz~\cite{MSidler2017}:
\begin{align}
    \ket{\Psi_\B p} &= a^\dagger_\B p \ket{\Phi} \label{eqn:ansatz} \\
    &= \left(\phi^c_\B p c^\dagger_\B k + \phi_\B p x^\dagger_\B p + \sum_{\B k \B q} \phi_{\B p \B k \B q} x^\dagger_{\B p + \B q - \B k} e^\dagger_\B k e_\B q  \right) \ket{\Phi} \nonumber
\end{align}
where $\ket{\Phi}$ denotes the ground state of the Fermi sea and the cavity. Note that we introduced the polaron-polariton creation operator $a^\dagger_\B p$, which obeys commutation relations that are approximately bosonic. The attractive polaron oscillator strength is given by $|\phi_\B p|^2$ and grows as $\sim E_{\mathrm{F}}$ for low doping $(E_{\mathrm{F}} < E_{\mathrm{T}})$. The oscillator strength of the bare exciton is shared among the two polaron branches, ensuring the coupling of both the attractive and repulsive branch to the cavity to form polaron-polaritons with a (reduced) normal mode splitting  (Fig.~\ref{fig: figure1}f).

The deviation of the exciton- upper and lower polariton resonance (UP and LP) lineshapes from a Lorentzian, depicted in Fig.~\ref{fig: figure1}e, stems predominantly from interference effects and indicate that polariton line broadening is primarily due to cavity losses~\cite{zeytinoglu18}. However, the excess broadening of the polaron-UP as compared to the polaron-LP hints towards the presence of residual polaron interactions. In the following section, we describe experiments that probe the transient changes in the transmission spectrum due to the evolution of a polaron subject to these residual interactions.

\begin{figure*}
            \includegraphics[width=\textwidth]{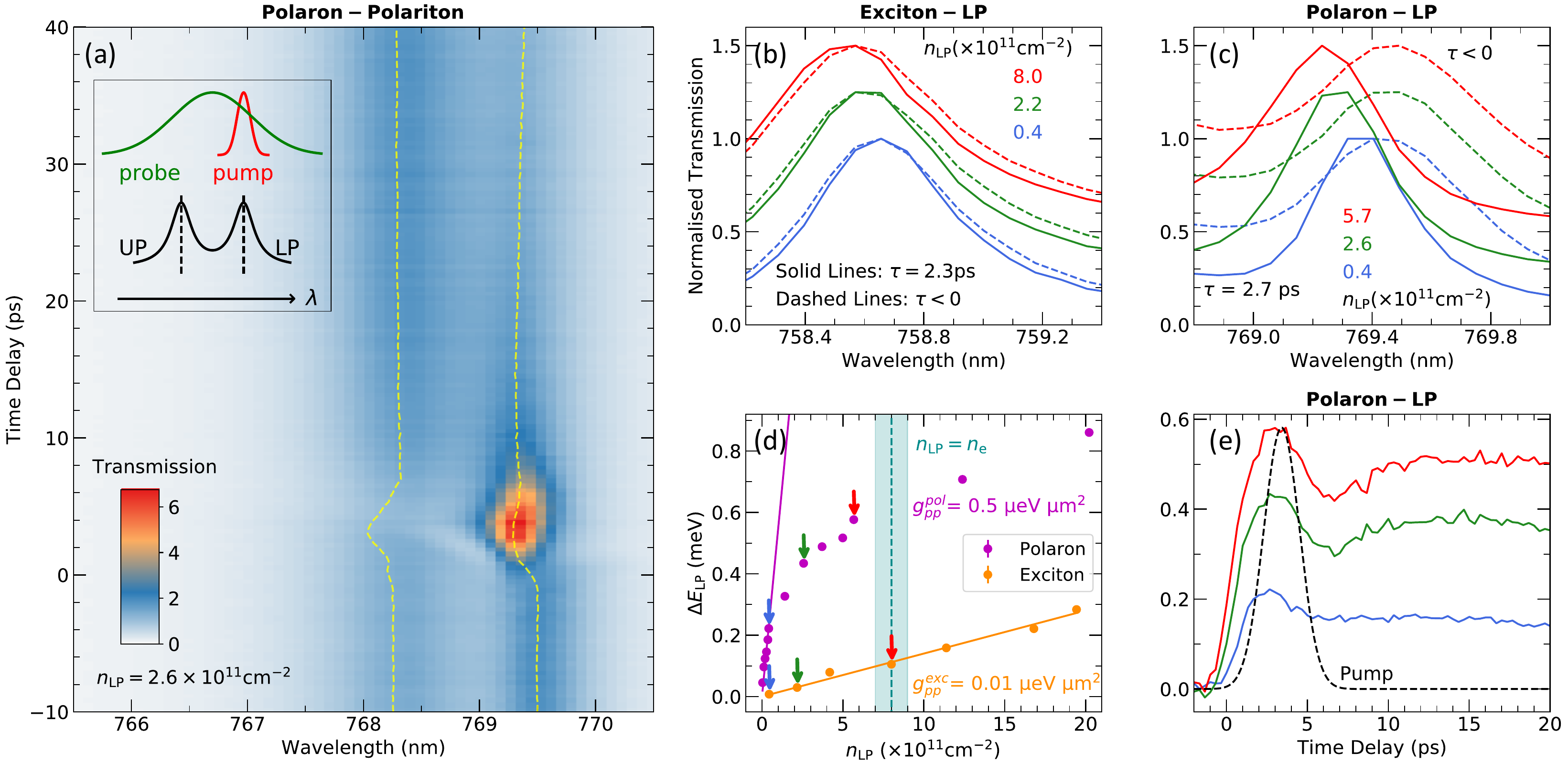}
    \caption{Blueshift of the lower polariton due to interactions. (a) A typical time delay scan of the probe transmission spectrum from a linear cross-polarised pump-probe experiment. In this case, the polaron lower-polariton (LP) is resonantly pumped. Inset: an illustration of the energy level scheme of the upper and lower polariton (UP and LP) branches and the pump and probe fields. The pump field is spectrally narrow and tuned in resonance to the LP while the probe field is spectrally broad and covers both branches. Yellow dashed lines: Evolution of UP and LP resonance wavelengths as a function of time delay. (b) Normalised probe transmission spectrum of the exciton-LP under resonant pumping for various polariton densities $n_{\mathrm{LP}} = 8, 2.1$ and $0.4 \times 10^{11} \mathrm{cm^{-2}}$. An offset has been added for clarity. Solid lines: The probe transmission spectrum measured at $\tau = 2.7\mathrm{ps}$. Dashed lines: The average $\tau < 0$ probe transmission spectrum obtained by taking the mean of the transmitted signal at a given wavelength over 5 different negative time delay values. (c) Analogous to (b) but for the polaron-LP at polariton densities $n_{\mathrm{LP}} = 5.7, 2.6$ and $0.4 \times 10^{11} \mathrm{cm^{-2}}$. (d) Energy shift of the exciton- and polaron-LP at $\tau = 2.7 \mathrm{ps}$ measured with respect to $\tau < 0$ as a function of pump power. Blue shaded area indicates the regime where polariton density and electron density are comparable in the system. The error bars are too small to be visible in the plot. For example, for the points indicated by the red arrows, the error bars are $4 \mu \mathrm{eV}$ and $9 \mu \mathrm{eV}$ for the energy shift of the polaron and the exciton respectively. (e) Evolution of the LP energy shift as a function of time delay. There is a contribution from coherent polaritons at short timescales and one from incoherent polarons at longer timescales. Black dashed line: indicates the arrival of the pump pulse as determined in an independent measurement (see Appendix B).}
\label{fig: figure2}
\end{figure*}

\section{Enhanced Polariton Nonlinearities due to Residual Quasiparticle Interactions}
\label{sec:three}
The time-resolved pump-probe spectroscopy setup is shown in Figure~\ref{fig: figure1}g. We use a spectrally narrow pump field ($1$~meV) with pulse duration of $\tau_{\mathrm{pump}} = 2.62 \pm 0.01$~ps to inject a majority polariton population up to as high as $n_\mathrm{LP} = 2 \times 10^{12} \mathrm{cm}^{-2}$. The probe field is spectrally broad ($12$~meV) with a pulse duration of $\tau_{\mathrm{probe}} = 0.3 \pm 0.1 $~ps and it injects a minority polariton population in a linear orthogonal polarisation unless stated otherwise. We monitor the pump-induced changes of the probe transmission as a function of time delay $\tau$. Zero time delay ($\tau = 0$) is defined with respect to the leading edge of the pump pulse, i.e., the probe pulse impinges on the sample concurrently with the leading edge of the pump. We expect the probe transmission (i) to be unperturbed by the pump for $\tau < 0$, (ii) to show a nonlinear response due to the presence of pump-injected coherent polaritons for $0 < \tau \lesssim (\tau_{\mathrm{pol}} + \tau_{\mathrm{pump}})$, and (iii) to be modified due to interactions with pump-induced longer-lifetime high-momentum (incoherent) polaron population for $\tau \gtrsim (\tau_{\mathrm{pol}} + \tau_{\mathrm{pump}})$.

Figure~\ref{fig: figure2}a shows a typical pump-probe measurement of the probe transmission as a function of $\tau$. Two distinctive features are observed: a strong blueshift of the polaron-LP energy and the amplification of the probe transmission. We discuss the former in this section and the latter in the next (Sec.~\ref{sec:four}). In the first set of experiments, we compare changes to the probe transmission in two cases: when the pump field is tuned into resonance with the exciton-LP ($n_{\mathrm{e}} = 0$) and the polaron- LP transition ($n_{\mathrm{e}} = 8 \times 10^{11} \mathrm{cm}^{-2}$, see Appendix D). The schematic is shown in the inset of Figure~\ref{fig: figure2}a. 

In Figure~\ref{fig: figure2}b and c, we show the normalised transmission spectra of the probe pulse, zoomed in onto the exciton-LP (polaron-LP) at $\tau < 0$ (see dashed lines) and $\tau = 2.3~\mathrm{ps}~(2.7~\mathrm{ps})$ (see solid lines), the time delay at which the maximum blueshift is observed. From this data, we extract $\Delta E_\mathrm{LP}$ -- the magnitude of the maximal LP resonance energy shift relative to its value for negative time delay. By monitoring this differential energy shift, we isolate the anharmonicity of the polariton resonance arising from interactions with the pump-field injected polaron-polaritons from cumulative changes in the optical response stemming from multiple pulse excitations that last for timescales exceeding the pulse repetition period. In Figure~\ref{fig: figure2}d, we plot $\Delta E_{\mathrm{LP}}$ as a function of the polariton density (see Appendix C for a description of the method used to determine the polariton density). We find a dramatic enhancement of the polariton-polariton interaction strength by a factor $\sim 50$ in the presence of itinerant electrons. Specifically, in the case of the polaron-LP, $g^{pol}_{\mathrm{pp}} = 0.5~\mathrm{ \mu eV \mu m}^{2}$ while for the exciton-LP, $g^{exc}_{\mathrm{pp}} = 0.01~\mathrm{\mu eV \mu m}^{2}$. Our experimentally measured value for $g^{pol}_{\mathrm{pp}}$ agrees with the theoretically expected value discussed in Section~\ref{sec:five}. The bare exciton-exciton interaction strength extracted from this result is $g^{exc}_{\mathrm{xx}} = 0.08~\mathrm{\mu eV \mu m}^{2}$, in agreement with earlier measurements~\cite{Scuri18-PRL, Barachati2018}.

For excitons, it is well-understood that at low densities, nonlinearities stem mainly from Coulomb interaction. The effective exciton-exciton interaction then scales with the exciton Bohr radius, $a_\mathrm{B}$, which is small in TMDs compared to GaAs-based semiconductors. In contrast, for excitons that form polarons in the presence of an electronic environment, the effective interaction has two main contributions: PSF stemming from the finite electron number inside the cavity area and the exchange of an electron-hole pair between two polarons. Here, we focus on the former which we find to be sufficient to explain our experimental observations. The second term of Eq.~\ref{eqn:ansatz} tells us the effective number of electron-hole pairs used from the Fermi sea to dress a single exciton impurity. As the number of impurities is increased, each successive exciton gets dressed by an increasingly depleted Fermi reservoir. This depletion of the electron reservoir has two main consequences: 1. Effective blue shift of the resonance as not as many electrons can participate in the dressing. 2. Reduction of the quasiparticle weight. The second effect shows up as a reduced oscillator strength that results in a reduction of the normal mode splitting as well as a blue shift of the LP. Consistent with the two described processes, we indeed experimentally observe both a blueshift of the polaron-LP and a reduced splitting.

We emphasise that the observed saturation of the energy shift for the polaron-polaritons as compared to the exciton-polaritons cannot be due to changes in the detuning between the lower polariton resonance and the pump laser. Rather, we attribute the saturation of the polaron-polariton energy shift to the breakdown of the polaron picture as the increasing exciton density becomes comparable to the electron density (indicated by the blue region of Fig~\ref{fig: figure2}d). For $n_{\mathrm{LP}} > n_{\mathrm{e}}$, $\Delta E_{\mathrm{LP}}$ continues to increase but with a gentler slope. In this limit, we are dealing with a Bose-Fermi mixture consisting of degenerate electrons and polaritons and we expect the simple Fermi-polaron model to be invalid.

We note that signatures of aforementioned cumulative long-time-scale changes on polariton spectra are visible in the data at $\tau < 0$ (dashed lines in Figure~\ref{fig: figure2}b). Remarkably, while the exciton-LP for $\tau < 0$  shows a significant amount of blueshift as the excitation power is increased, the polaron-LP resonance energy remains largely unchanged. The observed blueshift of the exciton resonance at $\tau < 0$ could originate from optical doping effects due to strong pump laser illumination, effectively turning the exciton into a repulsive polaron. 

In Figure~\ref{fig: figure2}e, we turn our attention to the time delay dependence of $\Delta E_\mathrm{{LP}}$. Two timescales can be seen to dominate the behaviour: the resonance first blueshifts monotonically as it follows the excitation pulse shape and reaches its peak at $\tau \sim 2.7 \mathrm{ps}$ before it starts to decay, consistent with the evolution of coherent polariton population resonantly injected by the pump. Then, a second mechanism which builds up over $\sim 12 \mathrm{ps}$ takes over that eventually decays over a much longer time. The buildup of the second peak grows noticeably more prominent as power is increased. We tentatively attribute this additional blueshift to the feeding of high-momentum polaron states from higher-energy optical excitations that are either generated by two-photon absorption or by Auger processes. Since high-momentum polarons contribute to PSF as well, they also lead to a blueshift of the polaron-LP resonance. In this scenario, the rise time of the blueshift should be determined by the feeding time from higher energy states populated by two-photon absorption and/or the Auger recombination rate, while the decay indicates the loss rate of these high-momentum polarons. For the exciton-LP, there is no clear evidence of an incoherent contribution which is likely due to the absence of itinerant carriers (see Appendix F).

\begin{figure}
            \includegraphics[width=\textwidth]{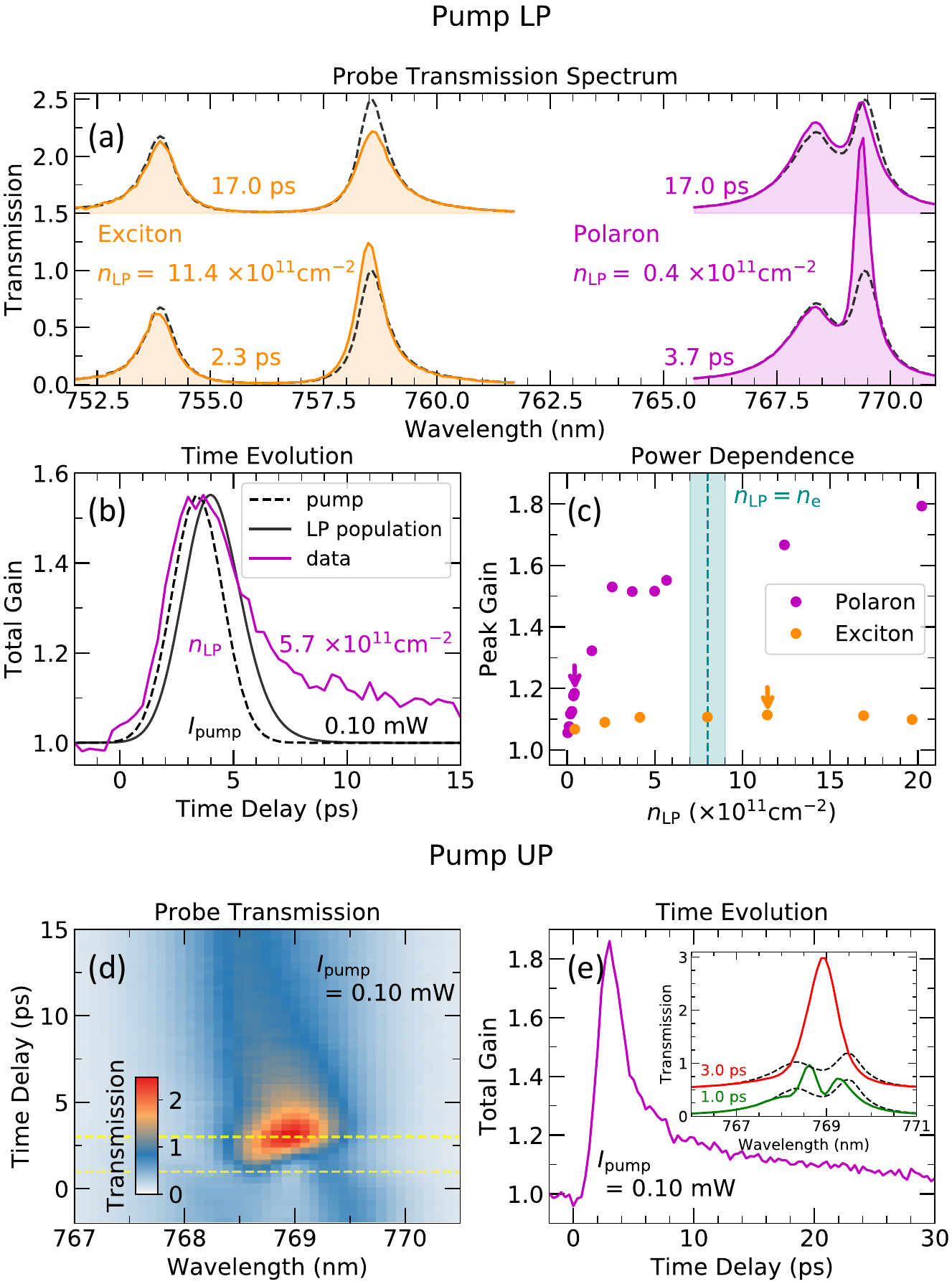}
    \caption{Polariton amplification observed in transmission. (a) Solid lines: The probe transmission spectrum under resonant pumping of the exciton lower-polariton (LP) and polaron-LP at different time delays, $\tau$. Dashed lines: The average $\tau < 0$ probe transmission spectrum obtained by taking the mean of the transmitted signal at a given wavelength over 5 different negative time delay values. (b) Evolution of the total gain for polaron-LP as a function of $\tau$. Gain for a given $\tau$ is calculated by dividing the integrated transmission under both UP and LP branches by the average integral taken at $\tau < 0$. Black dashed line: Indicates the arrival of the pump pulse as determined in an independent measurement (see Appendix B). Black solid line: The expected time evolution of the coherent LP population injected by the pump. (c) Dependence of the peak gain on the density of LPs injected by the pump. Blue shaded area: Indicates the regime where polariton density and electron density are comparable in the system. (d) Time delay scan of the probe transmission spectrum for resonant pumping of the polaron-UP. (e) Time evolution of integrated gain corresponding to (d). Inset: The probe transmission spectrum at $\tau = 1~\mathrm{ps}$ and $3~\mathrm{ps}$ indicated by yellow dashed lines in (d).}
\label{fig: figure3}
\end{figure}

\section{Polariton Gain}
\label{sec:four}
A second striking feature in Figure~\ref{fig: figure2}a is the enhancement of the polaron-LP transmission for $2$ ps $< \tau < 6$ ps as compared to $\tau < 0$ transmission. The magnitude of the increased transmission cannot be explained by a simple change in the cavity-polaron detuning as discussed in the previous section but rather suggests an amplification of polaritons. In Figure~\ref{fig: figure3}a, we show example linecuts for the exciton and polaron-polariton spectrum for two different $\tau$ contrasted with the typical spectrum at $\tau < 0$. We define the net transmission gain at every $\tau$ by the ratio of the integral under the shaded regions to the average integral at $\tau < 0$ (area under the dotted lines). A simple change in the cavity detuning would not lead to net gain deviating from $1$ since the cavity content of the two polariton branches must always be unity. Therefore net spectrally integrated gain exceeding unity suggests an amplification of the probe field. In semiconductor-microcavity systems, optical gain can be observed due to parametric scattering of polaritons when pumping conditions are fine-tuned such that pump, signal and idler conserve energy and momentum~\cite{baumberg00-PRL}. However, the significant gain we observe for the polaron-LP in Figure~\ref{fig: figure3}b considerably outlives the coherent LP population (black solid line), suggesting a contribution from a reservoir of long-lived incoherent polarons and rules out the possibility of a coherent process as the sole mechanism. 

There are a few possibilities as to how an incoherent polaron population can be generated. Firstly, due to the small normal mode splitting of the polaron-polariton, there is finite overlap of the LP with high-momentum polaron states which are not coupled to the cavity. Polarons can be created directly in high-momentum states due to the presence of disorder. Secondly, as we argued in the previous sub-section, Auger recombination or relaxation following direct two-photon absorption can efficiently generate high-momentum attractive polaron states which are immune to radiative decay. We find such processes to be consistent with the second rise of the blueshift in Figure~\ref{fig: figure2}e which grows in prominence relative to the first peak as pump power is increased. These high-momentum polarons can scatter electron-hole pairs from the Fermi sea of electrons to relax their momentum and energy into the polaron-LP state. Bose stimulation of this process due to a seed population from the probe then leads to gain. We verified that the gain factors observed for all pump powers are independent of the incident probe power.

In Figure~\ref{fig: figure3}c, we plot the power dependence of the peak gain for the exciton- and polaron-polariton. The saturation behaviour of the gain for the polaron-polariton resembles that of the blueshift: it first increases sharply and then continues with a gentler slope as $n_{\mathrm{LP}}$ approaches $n_{\mathrm{e}}$. The maximum gain for the polaron-polariton is $\sim 1.8$ while for the exciton-polariton, it remains constant at 1.1 for all $n_{\mathrm{LP}}$. We remark that the mechanism underlying weak exciton-polariton gain and its saturation behaviour remain unclear.

In order to understand the \textit{polaron}-LP gain mechanism, we studied gain under UP pumping since the population of high momentum polarons is expected be more efficient when pumping the UP. In Figure~\ref{fig: figure3}d, we plot the probe transmission spectrum at different $\tau$. Gain is observed to start from the UP wavelength and redshifts with $\tau$. However, as compared to pumping the LP, the polaron-polariton splitting collapses into a single amplified mode (inset of Figure~\ref{fig: figure3}e) which at even longer time delays, evolves into the UP resonance (Figure~\ref{fig: figure3}d). It is not obvious if this is due to a non-uniform gain spectrum or a breakdown of the strong coupling regime. The peak gain when pumping the UP was found to be $\sim1.8$ as compared to 1.5 when pumping the LP for the same pump power (Figure~\ref{fig: figure3}e).

\begin{figure}
            \includegraphics[width=\textwidth]{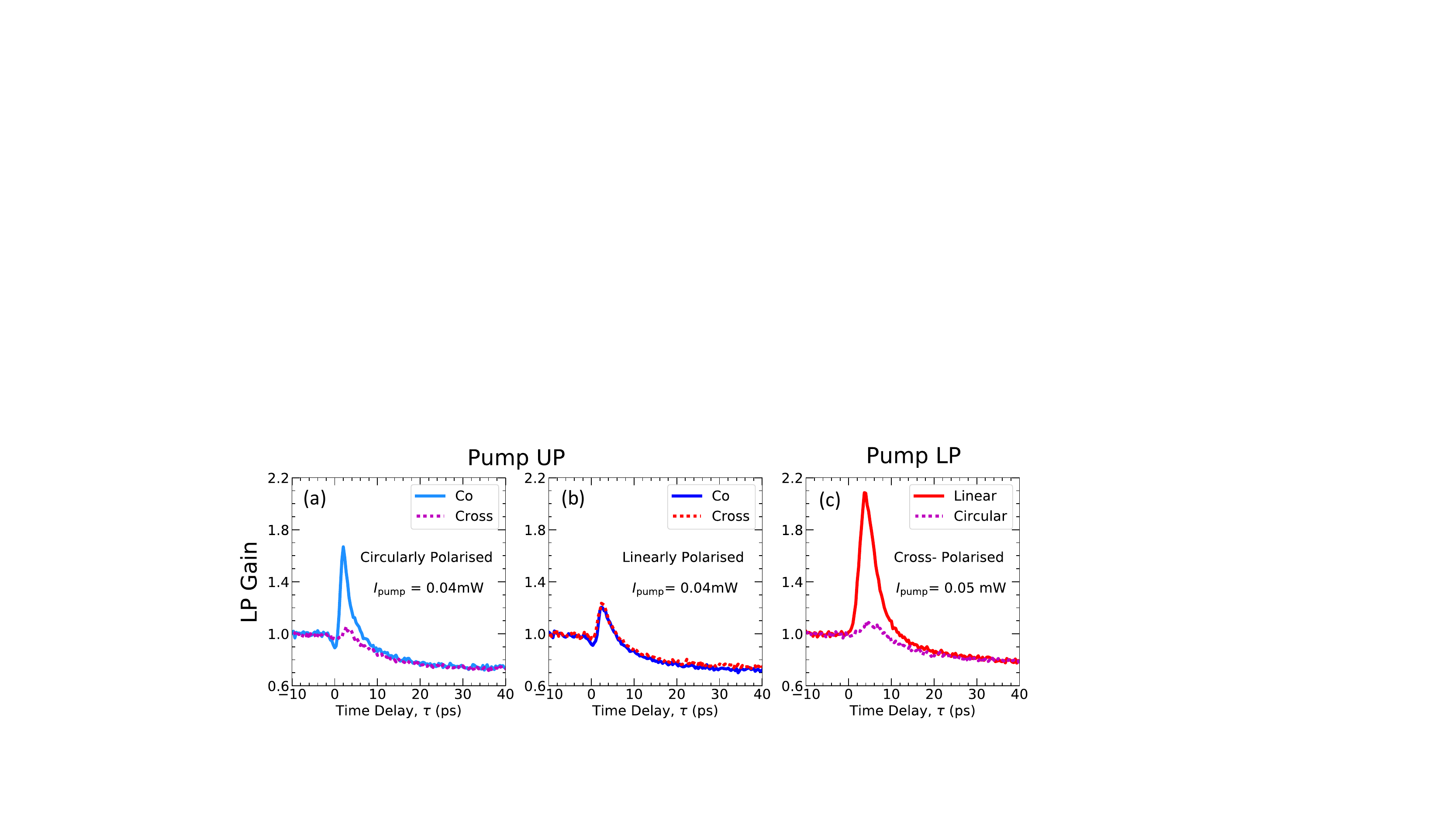}
    \caption{Evolution of lower polariton (LP) gain as a function of time delay $\tau$ for various pump-probe polarisation configurations. LP gain for a given $\tau$ is calculated by taking the ratio of the integral under the LP branch at $\tau$ to the average integral taken at $\tau <0$. (a) LP gain for co- or cross-polarised probe are plotted when resonantly pumping the polaron-UP with a circularly-polarised laser. (b) Analogous to (a) but for linear-polarisation. (c) The polaron-LP is resonantly pumped with either linearly or circularly polarised laser while the probe is always cross-polarised.}
\label{fig: figure4}
\end{figure}

Pumping the UP provides the advantage that both co-and cross polarised pump-probe experiments can be done since pump photons can be filtered spectrally instead of using polarisation-suppression. The polarisation DOF is important due to valley-dependent optical selection rules in monolayer TMDs: circularly-polarised photons excite polarons in a single valley which predominantly interact with electrons of the opposite valley. Therefore, we expect that pump-generated reservoir polarons can only be stimulated by probe-generated LPs in the same valley. To verify this, we pump the UP with circularly-polarised photons and probe with cross- and co- circularly polarised pulses. Due to spectral filtering, we limit the integration area to the LP branch when monitoring the gain. Consequently, changes in the cavity detuning can lead to gain values deviating from unity. Indeed for circular polarisation, the cross-polarised probe gain was found to be negligible as compared to its co-polarised counterpart (see Figure~\ref{fig: figure4}a). On the other hand, the co- and cross-polarised probe gain are identical in linear polarisation. This suggests that the gain process conserves the valley population but not the valley coherence, which is consistent with the proposed mechanism. The entanglement of polarons with the electrons during the scattering process leads to a loss of the phase relation between the valley populations. The polarisation dependence and the long-lived nature of the gain lends strong evidence for the significance of an incoherent but valley-preserving reservoir population. We also note that similar polarisation dependent behaviour is observed when pumping the LP as shown in Figure~\ref{fig: figure4}c.

\section{Theoretical Model}
\label{sec:five}
We now turn to the discussion of the theoretical model that underlies the interpretation of our experimental results. In order to model the observed phenomena theoretically, we use a wavefunction approach based on the Chevy ansatz given by Eq. (4). The interaction between the polarons described by this ansatz arises in our model from the symmetrisation of two polaron wavefunctions that accounts for the underlying fermionic nature of the polaron dressing cloud. Physically this accounts for the PSF and the related depletion of the electronic medium. As outlined below, in a similar fashion, one can calculate the residual interactions between polarons and electrons that lead to the decay of high-momentum polarons to the polaron-LP state.

Our experiments probe polariton-polariton interactions in the nonequilibrium limit where the radiative lifetime of optical excitations is comparable to or shorter than the electronic timescales. Since this is intrinsically a non-equilibrium problem, we have to consider interactions between polaron states that are not necessarily the lowest energy optically excited many-body state. While a full Keldysh Green’s functions approach is a method of choice to analyze such non-equilibrium problems in the presence of dissipation, we show here that a wavefunction approach leads to remarkably good agreement between theory and experiment.

\subsection{Fermi polaron- polaritons}\label{sec:polarons}
The coefficients of the polaron-polariton wavefunction given in Eq.~\eqref{eqn:ansatz}, $\phi_\B p^c$,  $\phi_\B p$ and $\phi_{\B p\B k \B q}$ are determined by the minimisation of $\bra{\Phi} a_\B p  (H-E_\B p) a_\B p^\dagger \ket{\Phi}$, where the polaron energy $E_\B p$ is a Lagrange multiplier ensuring the normalization of the wavefunction. Performing this minimization we obtain the following expression for the Fermi-polaron  quasiparticle energy: 
\begin{align}
E_\B p =& \nu_\B p + \frac{\Omega^2}{E_\B p-\omega_\B p - \Sigma_{xe}(\B p, E_\B p)}. 
\end{align}
The above equation shows that the hybridization with the excitonic degrees of freedom results in a self-energy for the photon. The exciton, in turn, acquires a self-energy $\Sigma_{xe}$ due to its interactions with the electron. These interactions are renormalized from the contact potential $v$ and determined by the electron-exciton scattering T-matrix, that accounts for effects of the finite electron density:
\begin{align}
    T(\B p, \omega)^{-1} = v^{-1} - \frac{1}{\cal A}\sum_{|\B k|>k_F} \frac{1}{\omega-\epsilon_{\B k} -\omega_{\B p-\B k}},
\end{align}
where $\B p$ and $\omega$ denote the total momentum and energy of the exciton and the electron. The self-energy is expressed in terms of the T-matrix as: 
\begin{align}
    \Sigma_{xe}(\B p ,\omega) =\frac{1}{\cal A}\sum_{|\B q|<k_F} T(\B p + \B q, \omega+ \epsilon_\B q).
\end{align}

The coefficients $\phi_\B p^c$, $\phi_\B p$ and $\phi_{\B p \B k \B q}$ can also be expressed in terms of the self-energy and the T-matrix: 
\begin{align}
    \phi_\B p^c &=\left(1 -  \frac{\partial}{\partial \omega} \left[\frac{\Omega^2}{\omega - \omega_\B p-\Sigma_{xe}(\B p ,\omega) }\right]_{\omega=E_\B p} \right)^{-1/2}\label{photon}\\
    \phi_\B p&=\left(1 -  \frac{\partial}{\partial \omega} \left[\frac{\Omega^2}{\omega - \nu_\B p} +\Sigma_{xe}(\B p ,\omega) \right]_{\omega=E_\B p} \right)^{-1/2}\label{exciton}\\
    \phi_{\B p \B k \B q}&=\frac{T(E_\B p + \epsilon_\B q, \B p+\B q)\theta(|\B k|-k_F)\theta(k_F-|\B q|)}{E_\B p - \omega_{\B p + \B q-\B k}-\epsilon_\B k + \epsilon_\B q} \frac{\phi_\B p}{\cal A}.\label{wavefunction}
\end{align}
The terms in the square brackets in Eqs. \eqref{photon} and \eqref{exciton} represent the photon and the exciton self-energy, respectively.

In the following, we will suppress the zero momentum label when discussing zero-momentum polarons, such that $\phi^c\equiv \phi^c_0$, $\phi\equiv \phi_0$ and $\phi_{\B k\B q}\equiv \phi_{0\B k\B q}$. 

\subsection{Residual interactions between polarons}\label{sec:interactions}

For a sufficiently strong laser pump pulse, a finite polaron density is generated and interactions between the quasiparticles become important. To qualitatively assess their strength, we focus here on two polarons of zero total momentum. We calculate the difference $U=E_2 - 2 E_0$ where $E_2$ is the energy of a system with two polarons of zero total momentum, while $E_0$ denotes the energy of a single impurity of zero momentum. To first approximation the two-polaron wave function address by the probe pulse is given by $a^\dagger_0 a^\dagger_0 \ket{\Phi}$ (properly normalized). Higher-order contributions to the interaction can appear from the coupling to higher-energy states such as $a^\dagger_{\B k\neq 0} a^\dagger_{-\B k} \ket{\Phi}$ as well as excited polaron states. Since these states have higher energy, they lead to attractive interactions within perturbation theory. The experimental evidence in our case shows that the interaction between polaron-polaritons is effectively repulsive, suggesting that first-order contributions dominate over higher-order interaction terms. 

In this approximation, the strength of interaction between two polaron-LPs is given by: 
\begin{align}
    \frac{U}{\cal A} = \frac{\bra{\Phi} a_0 a_0 H a_0^\dagger a_0^\dagger \ket{\Phi}}{\bra{\Phi} a_0 a_0 a_0^\dagger a_0^\dagger \ket{\Phi}} - 2 E_0. \label{10}
\end{align}
Here, the overlap $\bra{\Phi} a_0 a_0 a_0^\dagger a_0^\dagger \ket{\Phi}$ follows from Wick's theorem:
\begin{align}
\bra{\Psi} &a_0 a_0 a^\dagger_0 a^\dagger_0 \ket{\Psi} = 2(1+I_h +I_e +I_{he})\label{overlap}
\end{align}
where we use the normalization condition $|\phi|^2 + \sum_{\B k \B q} | \phi_{\B k \B q}|^2=1$ and introduce the exchange integrals: 
\begin{align}
I_{h}=&- \sum_{\B k \B k' \B q} |\phi_{\B k \B q}|^2 |\phi_{\B k' \B q}|^2,  \label{Ih}\\
I_{e}=&-\sum_{\B k \B q \B q'} |\phi_{\B k \B q}|^2 |\phi_{\B k \B q'}|^2,\\
I_{eh}=& \sum_{\B k \B q \B q' } |\phi_{\B k \B q}|^2 |\phi_{\B k +\B q ' - \B q, \B q'}|^2 .\label{Ihe}
\end{align}
The factor of $2$ in \eqref{overlap} comes from the bosonic nature of the exciton.  One can verify that, at low electron densities, the hole exchange contribution $I_h$ is larger than all the other contributions by a factor of $1/(k_F a_T)^2$ where $a_T$ is the trion Bohr radius. This is because of the much smaller phase space available for the hole, which is limited to momenta $|\B q| < k_F$, compared to the phase space of the electron, which is limited to momenta $|\B k| > |k_F| $. Thus we can restrict our calculation to the hole-exchange contribution in the following so that: 
\begin{align}
    \bra{\Psi} &a_0 a_0 a^\dagger_0 a^\dagger_0 \ket{\Psi} \approx 2(1+I_h). 
\end{align}

Substituting this back into \eqref{10}, we obtain: 
\begin{align}
    \frac{U}{\cal A}&\approx \frac{1}{2}\bra{\Phi} a_0 a_0 H a_0^\dagger a_0^\dagger \ket{\Phi}(1 - I_h) -2 E_0\\
    &\approx \frac{1}{2}\bra{\Phi} a_0 a_0 H a_0^\dagger a_0^\dagger \ket{\Phi} -2 E_0 -2 E_0 I_h, \label{u}
\end{align}
where we used the fact that $I_h \ll 1$ and in the second line we replaced $ \frac{1}{2}\bra{\Phi} a_0 a_0 H a_0^\dagger a_0^\dagger \ket{\Phi} \approx 2 E_0$ in the term proportional to $I_h$, since corrections to this approximation are of higher order. 

The remaining expectation value can be evaluated by applying Wick's theorem once again. In this calculation, the terms $-2 E_0$ and $-2 E_0 I_h$ cancel all the intra-polaron direct and exchange terms that appear in $\frac{1}{2}\bra{\Phi} a_0 a_0 H a_0^\dagger a_0^\dagger \ket{\Phi}$. Thus, in agreement with the expectation, only the inter-polaron interaction terms will contribute to the interaction $U$. Furthermore, the direct inter-polaron interaction is zero due to number conservation. Keeping only the hole-exchange terms yields:
\begin{align}
    &\frac{U}{\cal A} \approx -\frac{2v}{\cal A}  \left[ \sum_{\B k \B k' \B q}\phi^*_{\B k \B q} \phi |\phi_{\B k' \B q}|^2 \right. \\
    &+\left. \sum_{\B k \B q \B k' \B q'}\phi_{\B k \B q}^*\phi_{\B k'\B q'}^* \phi_{\B k \B q'} \phi_{\B k'+\B q - \B q', \B q} -\sum_{\B k \B k' \B q' \B q} \phi_{\B k \B q'}^*  \phi_{\B k \B q} |\phi_{\B k' \B q}|^2  \right]\nonumber
\end{align}
To understand the various terms in $U/{\cal A}$, it is helpful to recognise that a single polaron is formed from the superposition between a bare photon $c^\dagger_0 \ket{\Phi}$, an exciton $x_0^\dagger \ket{\Phi}$ and an exciton entangled with a neutral excitation in the Fermi sea $\sum_{\B k \B q} \phi_{\B k \B q} x^\dagger_{\B q-\B k} e^\dagger_\B k e_\B q \ket{\Phi}$. The strength of the  hybridisation between the latter two states relies on the exciton's ability to create electron-hole pair excitations in the Fermi sea. which necessarily depends on the number of electrons in the Fermi sea.

However, when the two excitons are close to each other, and one of the excitons is in the dressed state $\sum_{\B k \B q} \phi_{\B k \B q} x^\dagger_{\B q-\B k} e^\dagger_\B k e_\B q \ket{\Phi}$ while the other exciton is in the state $x_0^\dagger \ket{\Phi}$, then it will be more difficult for the second exciton to evolve into the state $\sum_{\B k \B q} \phi_{\B k \B q} x^\dagger_{\B q-\B k} e^\dagger_\B k e_\B q \ket{\Phi}$, because some of the electrons in the Fermi sea are already correlated with the first exciton. This fact is captured by the first term in the square brackets. On the other hand, if also the second exciton is in the state $\sum_{\B k \B q} \phi_{\B k \B q} x^\dagger_{\B q-\B k} e^\dagger_\B k e_\B q \ket{\Phi}$, this state will have slightly higher energy because of the Pauli blocking between the dressing clouds of the two excitons. This repulsion mechanism is captured by the last two terms, which contain the exchange corrections coming from the exciton-electron (second term) interaction and exciton-hole interaction (third term). 

One can check that the last two terms go to zero as $\Lambda \to \infty$. Indeed, since $|\phi_{\B k \B q}|\propto 1/|\B k|^2$ for large momenta, one can check that the first sum in the square brackets diverges logarithmically with $\Lambda$, while the others remain constant. Therefore, only the first term can compensate the logarithmic decrease of the interaction strength $v$ as $\Lambda \to \infty$ and the interaction is given by: 
\begin{align}
&\frac{U}{\cal A}\approx -\frac{2v}{\cal A}  \sum_{\B k \B q}\phi^*_{\B k \B q} \phi \sum_{\B k'} |\phi_{\B k' \B q}|^2.
\end{align}
This expression allows us to make the connection to the intuitive picture of phase space filling and the related polaron-induced depletion of the electronic environment. Indeed, the last summation in Eq. 20 quantifies the depletion of the Fermi sea, since $\sum_{\B k'} |\phi_{\B k' \B q}|^2 = 1-\bra{\Psi} a_0 c_\B q^\dagger c_\B q a^\dagger_0 \ket{\Psi}$. Therefore, $v \sum_{\B k'} |\phi_{\B k' \B q}|^2$ denotes the change in the amplitude for creating an electron-hole pair with momenta $\B {k, q}$ due to the presence of the attractive polaron at zero momentum, again, making explicit the effect of a depletion-induced interaction.  

We can rewrite the above interaction in terms of the T-matrix using the definition for the polaron wavefunction introduced in \eqref{wavefunction}. Using $\partial T^{-1}(\B p,\omega) /\partial \omega = \frac{1}{\cal A}\sum_{|\B k|>k_F} 1/(\omega-\omega_{\B p -\B k}-\epsilon_\B k)^2$ we obtain: 
\begin{align}
   \frac{U}{\cal A} = \frac{\phi^4}{\cal A} \sum_{|\B q|<k_F}  \left. \frac{\partial T^2(\B q,\omega)}{\partial \omega}\right|_{\omega=E_0 +\epsilon_{\B q}}
\end{align}
While this expression has the same form for both exciton-polarons and polaron-polaritons, the coupling to the cavity will modify the excitonic weight $\phi$ and the energy of the attractive polaron $E_0$, leading to induced interactions of different strength for polaron-polaritons. 

\begin{figure}
    \centering
    \includegraphics[width=0.95\textwidth]{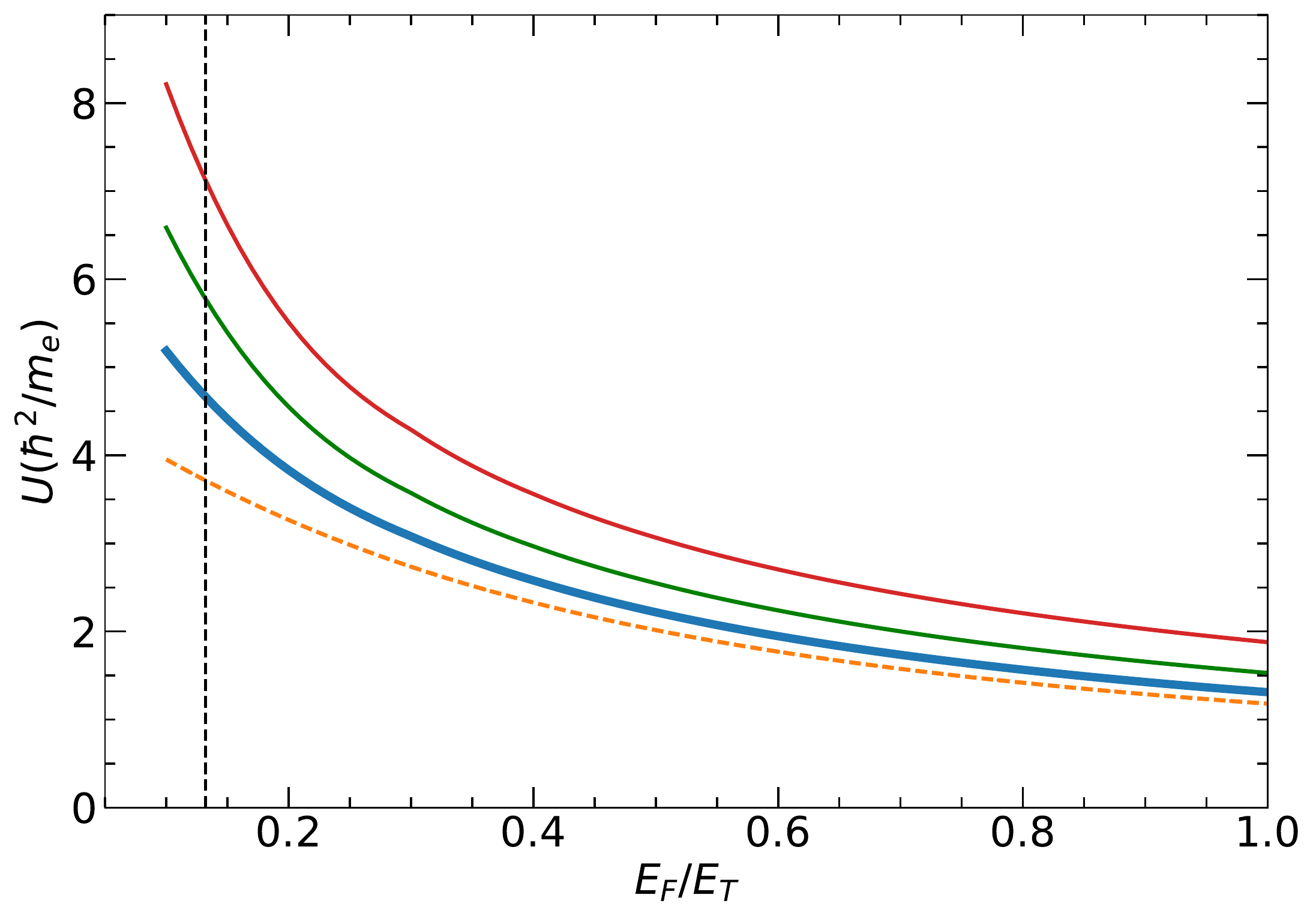}
    \caption{Residual interactions between polaron-polaritons and exciton-polarons. The thicker blue line denotes the interactions of Fermi polaron-polaritons as a function of Fermi energy when $\Omega=0.2 E_T$, corresponding to the experimental value. The green and red solid lines correspond to $\Omega=0.5 E_T$ and $\Omega=1 E_T$, respectively. For this plot we choose the detuning of the cavity  $\Delta=E_0 + \Omega^2/E_T$, where $E_0$ denotes the energy of the attractive exciton-polaron, to ensure that the polaron-polariton normal-mode splitting vanishes at zero Fermi energy. The orange dashed line represents the interaction strength of exciton-polarons in the absence of cavity coupling, divided by a factor of $4$. The enhancement of interactions with increasing cavity-polaron coupling $\Omega$ is evident. The black dashed line is a guide to the eye to illustrate the Fermi energy probed experimentally, i.e. $E_F/E_T=3.3/25$. Energies are in units of $\hbar^2/m_e$.}
    \label{fig:polaron-interactions}
\end{figure}

The resulting interaction strength is shown for parameters typical for our experiment in Fig.~\ref{fig:polaron-interactions}. As expected, polaron interactions increase as we decrease the Fermi energy, because they arise due to the Pauli-blocking between the Fermi-sea holes localized around the polarons. At the same time, we remark that interactions increase as the light-matter coupling $\Omega$ is increased suggesting that photon nonlinearities can be increased even further by reducing the cavity length. 

For a direct comparison to the experiment we choose the parameters $E_F=3.3$meV, $\Omega=5$meV, $E_T=25$meV. The cavity detuning in experiments is slightly different than the detuning chosen Fig.~\ref{fig:polaron-interactions}. To connect to the experiments we choose $\Delta=-25$meV, to ensure that the photonic weigth of the polaron-polariton agrees with the experimental value of $|\phi^c|^2=0.43$, For these parameters, our theory predicts a blue shift of $0.6 \mu \textrm{eV} \mu \textrm{m}^2 $ which agrees remarkably well with the experimental value of $0.5 \mu \textrm{eV} \mu \textrm{m}^2$. 

Before proceeding further, we note that in the equilibrium case the exchange of bosonic (electron-hole pair) excitations would provide an efficient mechanism to mediate polaron-polaron interactions \cite{yu2012induced,santamore2008fermion,hu2018attractive,camacho2018landau,tajima2018many,desalvo2019observation}. While such a generalization of the Ruderman-Kittel-Kasuya-Yosida (RKKY) interactions\cite{ruderman1954indirect,kasuya1956theory,yosida1957magnetic,desalvo2019observation} to the strong coupling regime provides a pathway to strong but attractive interactions, the low energy of these bosonic excitations ensures that the mediated interactions are strongly retarded and therefore ineffective in mediating interactions between short lived optical excitations. However, further analysis is necessary to incorporate the phase space filling effect into a general non-equilibrium theoretical framework.

\subsection{Polaron-electron interaction}\label{sec:electron-interactions}

As we argued in Sec.~IV, the observed probe gain originates from the residual interactions between polarons and the Fermi sea. For simplicity, we focus here on the gain in polaron-LP when polaron-UP branch is pumped. We envision that the relaxation of the excited polarons to polaron-LP state takes place in two sequential steps. In the first step, excitations in the polaron-UP branch decay into finite-momentum polarons by generating Fermi-sea electron-hole pairs, to form a long lived reservoir of finite-momentum polarons. In the second step, these polarons decay into the polaron-LP state by generating an additional electron-hole pair. This second process can be stimulated by a finite population in the final state, leading to net transmission gain for the probe field. Due to the complexity of the problem, we will only attempt to obtain an order-of-magnitude estimate of the gain, by calculating the rates for the two decay processes mentioned above.

The residual interactions between polarons and the Fermi sea can be estimated in an analogous way to the previous section.  To this end we evaluate the matrix element: 
\begin{align}
    &\bra{\Phi} e_\B q^\dagger e_\B k a_{\bf p+q-k}  H  a_{\B p}^\dagger \ket{\Phi}=\frac{S_{\B p \B k \B q}}{2}(E_{\B p}+E_{\B p+\B q-\B k} +\epsilon_\B k-\epsilon_\B q)\nonumber\\
    &+\frac{v}{\cal A} (\phi_\B p \phi_{\B p +\B q -\B k} +\sum_{\B k' \B q'} \phi_{\B p \B k' \B q'} \phi_{\B p+\B q -\B k,\B k', \B q'} )\nonumber\\
    &-\frac{v}{\cal A} \left( \sum_{\B k'\B q'} \phi_{\B p \B k'\B q} \phi_{\B p+\B q - \B k,\B k'\B q'}+  \sum_{\B k' \B q'} \phi_{\B p \B k \B q'} \phi_{\B p+\B q - \B k,\B k'\B q'}\right. \nonumber\\
    &\quad \quad  -\frac{1}{2}\sum_{\B k' \B q'} \phi_{\B p \B k \B q} \phi_{\B p+\B q - \B k,\B k'\B q'} - \frac{1}{2}\sum_{\B k'} \phi_{\B p \B k' \B q} \phi_{\B p +\B q -\B k} \nonumber\\
    &\quad \quad- \left. \frac{1}{2}\sum_{\B q'} \phi_{\B p \B k \B q'} \phi_{\B p +\B q -\B k}\right).
\end{align}
In the above, the term proportional to $S_{\B p \B k \B q} \equiv \bra{\Phi} e_\B q^\dagger e_\B k a_{\bf p+q-k}   a_{\B p}^\dagger \ket{\Phi}$ arises because we are working with a non-orthogonal basis and does not represent an interaction term. From the other terms, we focus only on the ones that remain finite in the UV limit $\Lambda \to \infty$, i.e. the 2nd, 3rd and 4th terms in the last parenthesis. In the low doping limit $k_F a_T \ll 1 $, relevant for  the description of our experiment the forth term dominates by a factor of $1/(k_F a_T)$, so we keep only this term, denoted by $V_{\B p\B k\B q}$ in the following analysis.  It is given by: 
\begin{align}
   V_{\B p \B k \B q} &\approx \frac{v}{2}\sum_{\B k'} \phi_{\B p \B k' \B q} \phi_{\B p +\B q -\B k}\\
    &=  \phi_\B p \phi_{\B p+\B q-\B k} T(\B p +\B q,E_\B p +\epsilon_\B q),
\end{align}
which agrees with previous calculations using Green's functions \cite{cotlet2018transport}. This matrix element describes the amplitude for the emission of an electron-hole pair with momenta $\B k$ and $\B q$ by a polaron of momentum $\B p$. We emphasize that the above matrix element can also describe the residual interaction between the upper polaron-polariton and the Fermi sea, if we replace the coefficients $\phi$ and energy $E_\B p$ with the corresponding values for the upper polaron-polariton.

To calculate the decay rate from the upper polaron-polariton into finite momentum polaron states we use Fermi's Golden rule: 
\begin{align}
    \Gamma^{U} =& \sum_\B p \Gamma^U_\B p \equiv \sum_\B p 2 \pi \sum_{\B q} \left|\frac{V^U_{0,\B q-\B p,\B q}}{{\cal A}^2}\right|^2\\
    &\delta(E_0^U-E_{\B p}-\epsilon_{\B q-\B p}+\epsilon_\B q) \theta(k_F-|\B q|)\theta(|\B q-\B p|-k_F)\nonumber
\end{align}
where $E_0^U$ denotes the energy of the excited upper polaron-polariton state and we used the superscript $U$ to emphasize that the interaction $V$ should be evaluated using the coefficients and the energy of the upper polaron-polariton. In the above $\Gamma_\B p$ denotes the decay rate into a the polaron state of momentum $\B p$ by emission of electron-hole pairs of total momentum $-\B p$. Choosing the same parameters as in the previous section we obtain $\hbar \Gamma^U\approx 2$meV. Since the radiative lifetime of polaritons is about $\hbar/(1 \textrm{meV})$ we conclude that pumping the upper polariton will create a sizeable reservoir of finite momentum Fermi-polarons. We remark that $\Gamma_\B p$ is strongly peaked at momenta $|\B p| \sim k_F/2.5$, implying that most polarons in the reservoir will have this momentum. 

We can also calculate the decay-rate from a finite-momentum polaron into a lower-polaron-polariton:
\begin{align}
    \Gamma^L_{\B p}= 2 \pi \sum_{\B q} \left|\frac{V_{\B p,\B p+\B q,\B q}}{{\cal A}^2}\right|^2 \delta(E_0-E_{\B p}-\epsilon_{\B p+\B q}+\epsilon_\B q)
\end{align}
Evaluating the above we obtain $\Gamma^L_{k_F/2.5} \approx  1\mu$eV$\mu$m$^2/{\cal A}$. Assuming that the finite-momentum polaron reservoir has a population density  between $n_0=10^{11}$cm$^{-2}$ and $n_0=10^{12}$cm$^{-2}$, we obtain a decay rate  $\hbar \Gamma^L \approx N_0 \Gamma^L_{k_F/2.5} =  1\mu$eV$\mu$m$^2 \times n_0$ which gives a value between $1$ and $10$meV. Assuming a polariton lifetime of $\hbar/1$meV, our simple calculation would therefore predict a net gain at polaron-LP exceeding unity, which agrees well with the experimentally observed gain. 

\section{Conclusion}
\label{sec:six}
In this work, we have explored polariton-polariton interactions and Bose-enhanced scattering of polaron-polaritons in a charge-tunable MoSe$_2$ monolayer embedded in a zero-dimensional optical cavity. We found polariton-polariton interactions to be enhanced by a factor of 50 when the monolayer is electron-doped as compared to the charge-neutral regime. This dramatic enhancement originates from the restriction of the oscillator strength of polarons formed within an optical spot with a fixed electron density.

Intuitively, this enhancement is a consequence of the rearrangement of the polaron dressing cloud to accomodate a larger number of optically injected impurities. In an undoped semicondutor, exciton-exciton interaction scales linearly with the Bohr radius ($a_B$), implying that strong light-matter coupling ($\Omega  \propto 1/a_B$) is normally associated with weak nonlinearity. In stark contrast, the new mechanism we identify leads to stronger polariton interactions for stronger light-matter coupling (Fig.~4). This dependence suggests that, against expectations, strong exciton binding may even help to increase polariton-polariton interactions in electron-doped samples. 

Our work further highlights the importance of time resolution when observing interaction-induced effects on polariton spectra. In particular, the measurements revealed spectral features that survive up to timescales far outliving the coherent polaritons. This observation sheds light on the possible existence of an incoherent reservoir which contributes to the blueshift of the polaron-LP resonance that exists for timescales $> 10$~ps. Residual polaron interactions also provide an efficient relaxation channel from these high-momentum reservoir states into the lower polariton mode, which when stimulated by probe-field injected polaritons, results in optical gain. Our findings suggest that injection of itinerant electrons into a monolayer TMD could overcome the relaxation bottleneck and enable the realization of a polariton laser~\cite{imamoglu96-PRA,schneider18-NatureCom}.  

The signatures revealed by nonlinear spectroscopy in our work represent an exciting new realm of polaron physics: prior results in this field are often well-described by the Chevy ansatz which models many-body dressing in terms of a single electron-hole pair. However, an accurate description of interactions that arise from the re-organisation of impurity screening by the bath will require a description of higher-order exciton-electron correlations.

Furthermore, our experiments have also made a first venture into the exciting regime of degenerate Bose-Fermi mixtures where the polariton density becomes comparable to the electron density. Addressing open questions such as the electron density dependence of the onset as well as magnitude of the saturation behavior in the observed blueshift and gain will provide insight to guide theoretical understanding. While such investigations are limited in the current sample due to the normal mode splitting of the polaron-polaritons rapidly diminishing when electron density is tuned, enhancing the cavity quality factor from $\sim1000$ to $10,000$ should allow a wider range of electron densities to be explored.

The data that support the findings of this Letter are available in the ETH Research Collection.
\begin{acknowledgments}
The authors gratefully acknowledge many enlightening discussions with Eugene Demler that shaped our understanding of polaron physics and its relevance for 2D materials. They also thank Jacqueline Bloch, Georg Bruun, Jerome Faist, Meinrad Sidler, Sina Zeytinoglu, Thibault Chervy and Ido Schwartz for useful discussions. This work is supported by a~European Research Council (ERC) Advanced investigator grant (POLTDES), and by a~grant from Swiss National Science Foundation (SNSF).
\end{acknowledgments}
\clearpage
%%%%%%%%%%%%%%%%%%
%SUPPLEMENTARY
%%%%%%%%%%%%%%%%%%
\appendix
\setcounter{equation}{0}
\setcounter{figure}{0}
\setcounter{table}{0}
\setcounter{page}{1}
\makeatletter
\renewcommand{\theequation}{S\arabic{equation}}
\renewcommand{\thefigure}{S\arabic{figure}}

\section{Sample Fabrication and Setup}
The monolayer MoSe$_2$, graphene and two h-BN flakes were mechanically exfoliated onto SiO$_{2}$ substrates. Then, they were stacked by picking up the graphene, top h-BN, MoSe$_2$, and the bottom h-BN, in that order, using a polycarbonate sacrificial layer on a PDMS stamp. The stack is deposited onto a fused silica substrate with a distributed Bragg reflector (DBR) coating: Ten alternating layers of NB$_2$O$_5$ and SiO$_2$. This constitutes the planar mirror of the 0-dimensional cavity. The DBR was designed to provide $>99.3\%$ reflectivity between 680-800nm with an intensity maximum at the DBR surface. The MoSe$_2$ and graphene flakes were contacted with Au on a 5nm Ti sticking layer. The concave mirror was prepared by ablating a dimple of radius of curvature 30 $\mathrm{\mu}$m onto a single mode fibre facet which was then coated with an identical DBR as the flat substrate. After fabrication of the sample, the graphene layer was found to be torn and could not provide reliable doping. Thus, an additional graphene layer was placed on top which restored gate-tunability in the sample.

The planar mirror substrate is mounted on 2 nano-positioners which provide in-plane spatial degrees of freedom. The fibre mirror is mounted on 1 nano-positioner which provides an out-of-plane degree of freedom.

\section{Optical Measurements}
The optical setup is illustrated in Fig.~\ref{fig: figure1}g of main text. The sample sits in a vacuum-pumped environment which is filled with 20 mbar of He exchange gas at room temperature. It is then immersed in a liquid He bath at 4.2 K for all optical measurements.

\begin{figure}
            \includegraphics[width=\textwidth]{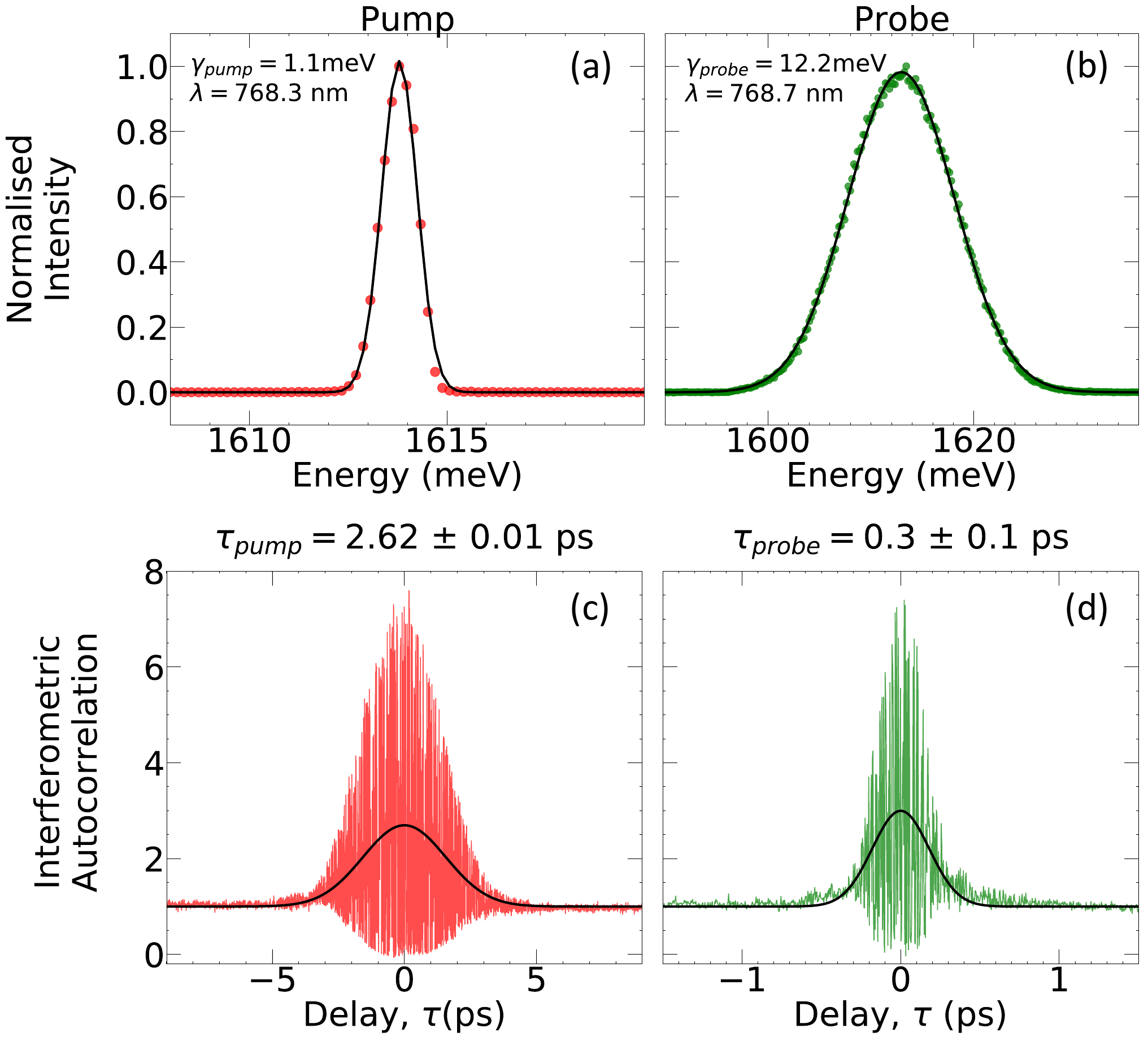}
    \caption{(a) and (b) A typical spectrum of the pump and probe pulse respectively after travelling through the optical fibres. (c) and (d) Interferometric autocorrelation signal of the pump and probe respectively. Black lines: Gaussian fits to the data from which the linewidths ((a) and (b)) and pulse duration ((c) and (d)) are extracted.}
\label{fig: pulseduration}
\end{figure}

\textit{Pulse preparation-} The output from a Ti:Sapphire femtosecond pulsed laser with 76 MHz repetition rate is split into two arms: pump and probe. The pump arm is spectrally filtered with a 4$f$ pulse shaping setup with a single grating. The 12 meV broad pulse is first spectrally dispersed using a transmission grating, and its spectrum selected with an adjustable slit aperture in front of a mirror and then recombined using the same grating. The probe pulse is spectrally unfiltered. Its optical path length difference with respect to the pump pulse is controlled by adjusting the position of a retroreflector. In this way, the probe can be made to arrive with a variable time delay, $\tau$ with respect to the pump. Both pump and probe pulses are then guided by optical fibres to the excitation arm of the transmission microscope. To avoid unwanted nonlinear effects in the fibres, we attenuate the laser powers to $<$ 1 mW before coupling them into the fibres. We check the spectrum of the pulse after travelling through the fibre in order to ensure such nonlinear effects are not present for the powers we are interested in (Fig~\ref{fig: pulseduration} a and b). In addition, there are also unwanted linear effects such as group velocity dispersion (GVD). Since the probe pulse is $\sim 12$ meV broad, it is affected more significantly than the pump pulse ($\sim 1$ meV). We compensate for the dispersion using a single grating pulse compressor. The pump and probe pulse durations are then measured using an interferometric autocorrelation setup in collinear geometry (Fig~\ref{fig: pulseduration} c and d). The uncertainty of the pulse durations arise from the possible deviations of the amount of dispersive elements (i.e, fibre lengths) incorporated in the autocorrelation measurement setup from the pump-probe experimental setup even as this was already taken into consideration in designing the former. We note that while we achieved significant compression of the probe, it remains not transform-limited.

\textit{Pump-probe measurement-} After the pulses are prepared, the pump and probe are coupled into the cavity via the free-space accessible side and the transmission is detected through the fibre. Using fibre polarization controllers and a polarizing-beamsplitter (PBS) in the detection setup, we can project the signal onto any two orthogonal polarizations. In a typical cross-polarized pump-probe measurement scheme, we measure the signal from the PBS arm that is cross-polarized with respect to the pump field using the spectrometer and as a function of $\tau$, we obtain the following: (i) the transmission spectrum when only the pump is turned on and (ii) the transmission spectrum when both pump and probe are turned on. In (i), we measure the cross-polarized polariton emission induced exclusively by the pump field and subtract this from (ii) in order to investigate how the transmission of the probe is influenced in the presence of the pump and as a function of $\tau$.

\textit{Measurement of the zero-delay between pump and probe-} After mixing the pump and probe pulses with a beamsplitter, the output from one port is sent to the cavity sample and the output from the second port is sent to a photodetector which measures the interfered signal. We note that the pump and probe travel through a common path after the beamsplitter and therefore, we can determine the zero delay between them as the point about which the interfered signal is symmetrical. Since this measurement is not performed in-situ but typically done right before or after the relevant pump-probe measurements, we cannot rule out the possibility of slight drifts in the true zero-delay position over time.

\section{Polariton Density Estimation}
We measure the reflection contrast of the bare cavity mode of $\eta \sim0.13$ between resonant and off-resonant conditions. This gives an estimate of the efficiency of the coupling into the cavity. The density of intracavity polaritons per pulse is then given by $\frac{\eta I_{pump} \varepsilon}{A \hbar \omega f_{rep}}$ where $f_{rep}$ is the repetition rate of the pulse train, $\varepsilon$ is the spectral overlap of the pump and the lower polariton branch, $A$ is the area of the excitation spot and $\hbar \omega$ is the photon energy. We are interested in the polariton density within the polariton lifetime $\tau_{pol}$ which can be written as $\frac{\eta I_{pump} \varepsilon}{A \hbar \omega f_{rep}} \cdot \frac{\tau_{pol}}{\tau_{pulse}+\tau_{pol}}$.

The spectral overlap $\varepsilon(\Delta_{\mathrm{cav}})$ is in fact a function of the detuning between the cavity and the polaron (or exciton) which determines the cavity content, $C$, and the resonance of the lower polariton (or the mode being pumped), $E_{\mathrm{LP}}$. Therefore, it is important to take into account the changes in $\varepsilon$ due to blueshifts of the polaron resonance when calculating the polariton density. We want to compute $\varepsilon$ during the pump pulse illumination, ie at $\tau_{\mathrm{max}}$ instead of at $\tau < 0$. However, when there is significant gain, it becomes difficult to determine the cavity content of the lower polariton from the area of the transmitted signal. The method we use is the following:

(1) From the polariton transmission data at $\tau < 0$ where there is no gain, we extract $C$ and $E_{\mathrm{LP}}$ and calculate $E_{\mathrm{cav}}$ using the experimentally determined value of $\hbar \Omega$ and the expression for the Hopfield coefficient:
\begin{equation}
    \label{eqn:Hopfield}
    |C|^{2} = \frac{1}{1 + \left( \frac{E_{\mathrm{LP}} - E_{\mathrm{cav}}}{\hbar \Omega} \right)^2}
\end{equation}
where $\hbar \Omega$ is the oscillator strength of the polaron.

(2) Then, we extract $E_{\mathrm{UP}}$ and $E_{\mathrm{LP}}$ from the $\tau_{\mathrm{max}}$ data where gain is observed. By taking the sum and difference of the two quantities and assuming that the cavity length remains unchanged when excited by the pump, we can find the altered oscillator strength $\hbar \Omega '$ and  $E'_{\mathrm{pol}}$ using
\begin{equation}
    E_{\mathrm{UP, LP}} = \frac{E_{\mathrm{cav}} + E_{\mathrm{pol}}}{2} \pm \frac{1}{2} \sqrt{(E_{\mathrm{cav}} - E_{\mathrm{pol}})^2 + 4|\hbar \Omega|^2} 
\end{equation}

(3) Then, the cavity content at $\tau_{\mathrm{max}}$ can be calculated using equation~\ref{eqn:Hopfield}.

\section{Electron Density Estimation}
We use a capacitive model to estimate the electron density. The capacitance per unit area between the top gate and the sample is given by:

\begin{equation}
    \frac{C}{A} =\left(\frac{t}{\epsilon_0 \epsilon_{hBN}} + \frac{1}{e^2 D(E_F)}\right)^{-1}
\end{equation}

where the first and second terms are the geometric and quantum capacitances respectively. $t = (88 \pm 5)$nm is the thickness of the h-BN flake, $\epsilon_{hBN} = 3.5 \pm 0.5$ is the static dielectric constant of h-BN, $m^* = 0.5 m_e$ is the effective electron mass in the conduction band and $D(E_{\mathrm{F}})$ is the density of electronic states at Fermi energy $E_F$. For $E_{\mathrm{F}} > 0$, one can neglect the quantum capacitance and write the Fermi energy as a function of applied gate voltage, $V_g$, as

\begin{equation}
    E_{\mathrm{F}}(V_g) = \frac{\pi \hbar^2 \epsilon_{hBN} \epsilon_0}{t e m^*} (V_0 - V_g).
\end{equation}

\begin{figure}
            \includegraphics[width=\textwidth]{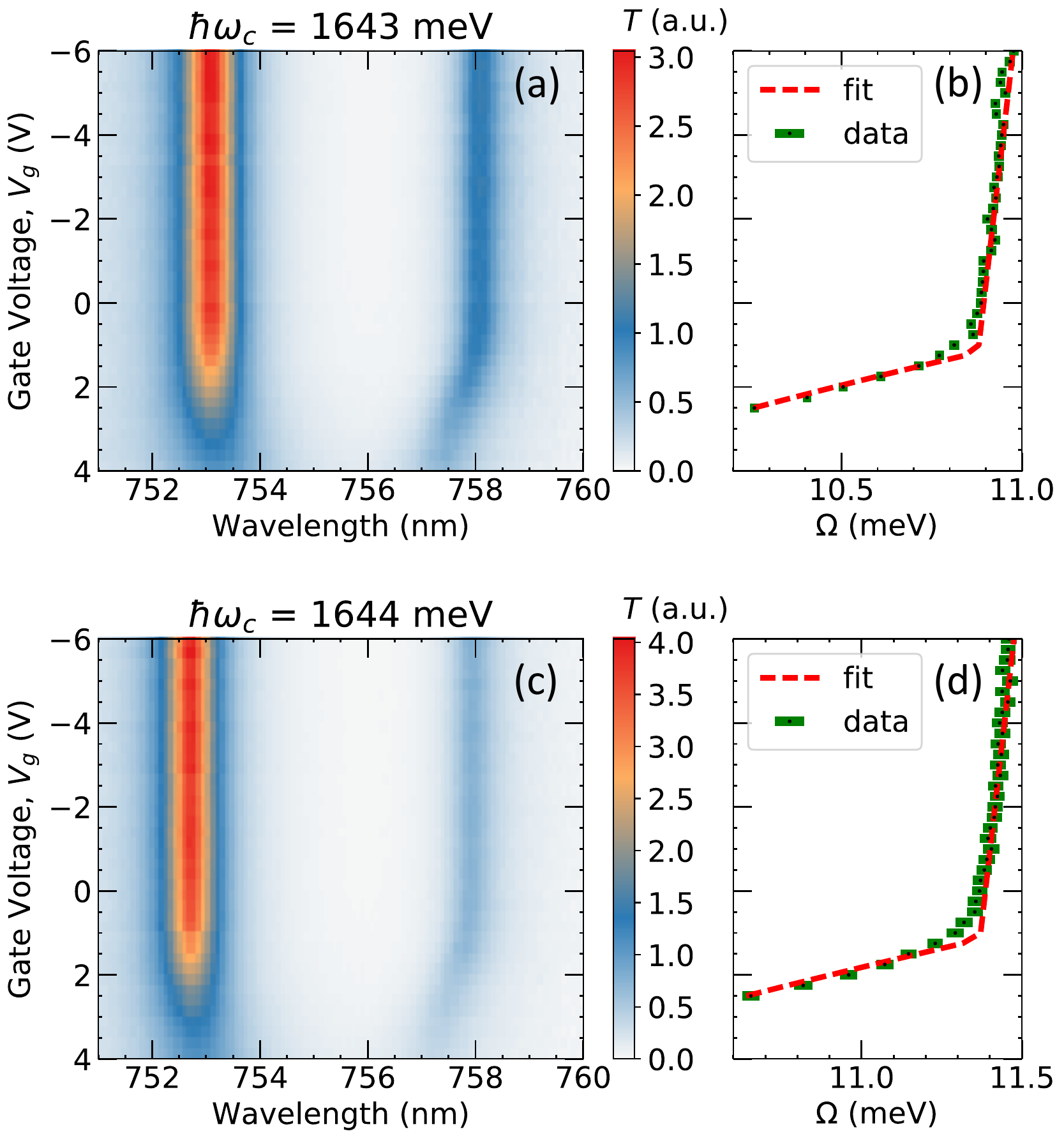}
    \caption{(a) Transmission $T$ spectrum of the repulsive-polaron polariton as a function of $V_g$ at a bare cavity mode energy of $\hbar \omega_c$ = 1643 meV. (b) Black dots with green error bars: Rabi splitting extracted from Lorentzian fits to the transmission spectrum. The red dashed line is a fit of the model detailed in the text to the data. (c), (d) Analogous to (a), (b) for $\hbar \omega_c$ = 1644 meV.}
\label{fig: electrondensity}
\end{figure}

In order to calculate the Fermi energy $E_{\mathrm{F}}$ and therefore the electron density $n_e$ from the applied gate voltage $V_g$, we first need to determine the voltage, $V_0$, at which we begin to dope the monolayer with itinerant electrons. To that end, we measured the transmission spectrum of the repulsive polaron-polariton as a function of gate voltage $V_g$. For small $E_{\mathrm{F}}$ (i.e. $V_g<V_0$), the Rabi splitting $\Omega_{\mathrm{rep}}$ is given by:

\begin{equation}
   \Omega_{\mathrm{rep}}(E_{\mathrm{F}}) = \sqrt{[\omega_{\mathrm{rep}}(E_{\mathrm{F}})-\omega_{\mathrm{c}}]^2 + [g_{\mathrm{rep}}(E_{\mathrm{F}})]^2},
   \label{eqn:rabi}
\end{equation}

where $\omega_{\mathrm{rep}}(E_{\mathrm{F}}) = \omega_x+ \beta E_{\mathrm{F}}$ is the energy of the repulsive exciton-polaron and $\omega_{\mathrm{x}}$ is the exciton energy; the second term accounts for the blueshift due to the presence of the Fermi sea and $\beta$ was previously found to be 0.8~\cite{MSidler2017}. $g_{\mathrm{rep}}(E_{\mathrm{F}})$ is the oscillator strength of the repulsive polaron which is $\sim g_0 (1 - \frac{1}{2} \frac{E_{\mathrm{F}}}{E_{\mathrm{T}}})$ for small $E_{\mathrm{F}}$, $E_{\mathrm{T}}$ refers to the trion binding energy which we take to be $25$ meV.

Figures~\ref{fig: electrondensity} (a) and (c) show the measured transmission spectrum of the repulsive-polaron polariton for two different cavity detunings as a function of $V_g$. It is observed (shown in Figures~\ref{fig: electrondensity} (b) and (d) that there are two distinct regimes for the Rabi splitting $\Omega_{\mathrm{rep}}$. $\Omega_{\mathrm{rep}}$ reacts much less sensitively to $V_g$ when increasing it from -6V to $\sim 1$V after which it starts to decrease sharply. We attribute this apparent slow increase of $E_F$ as a filling of localised states located in the midgap region resulting in a slight decrease of $\Omega_{\mathrm{rep}}$ which we can represent with a heuristic linear model. On the other hand, the behaviour of $\Omega_{\mathrm{rep}}$ is governed by equation~\ref{eqn:rabi} as soon as itinerant electrons start to populate the conduction band (for $V_g > V_0$). In our fits, all parameters in equation~\ref{eqn:rabi} were fixed except for $V_0$ which remained a fit parameter. We determine $V_0$ to be 1V. This implies that $n_e = (8 \pm 1) \times 10^{11}$ cm$^{-2}$ at $V_g = 5$V. We note that the uncertainty is dominated by that of $\epsilon_{hBN}$.

\section{Time Delay Dependence of Exciton-Polariton Blueshift}
\begin{figure}
            \includegraphics[width=0.7\textwidth]{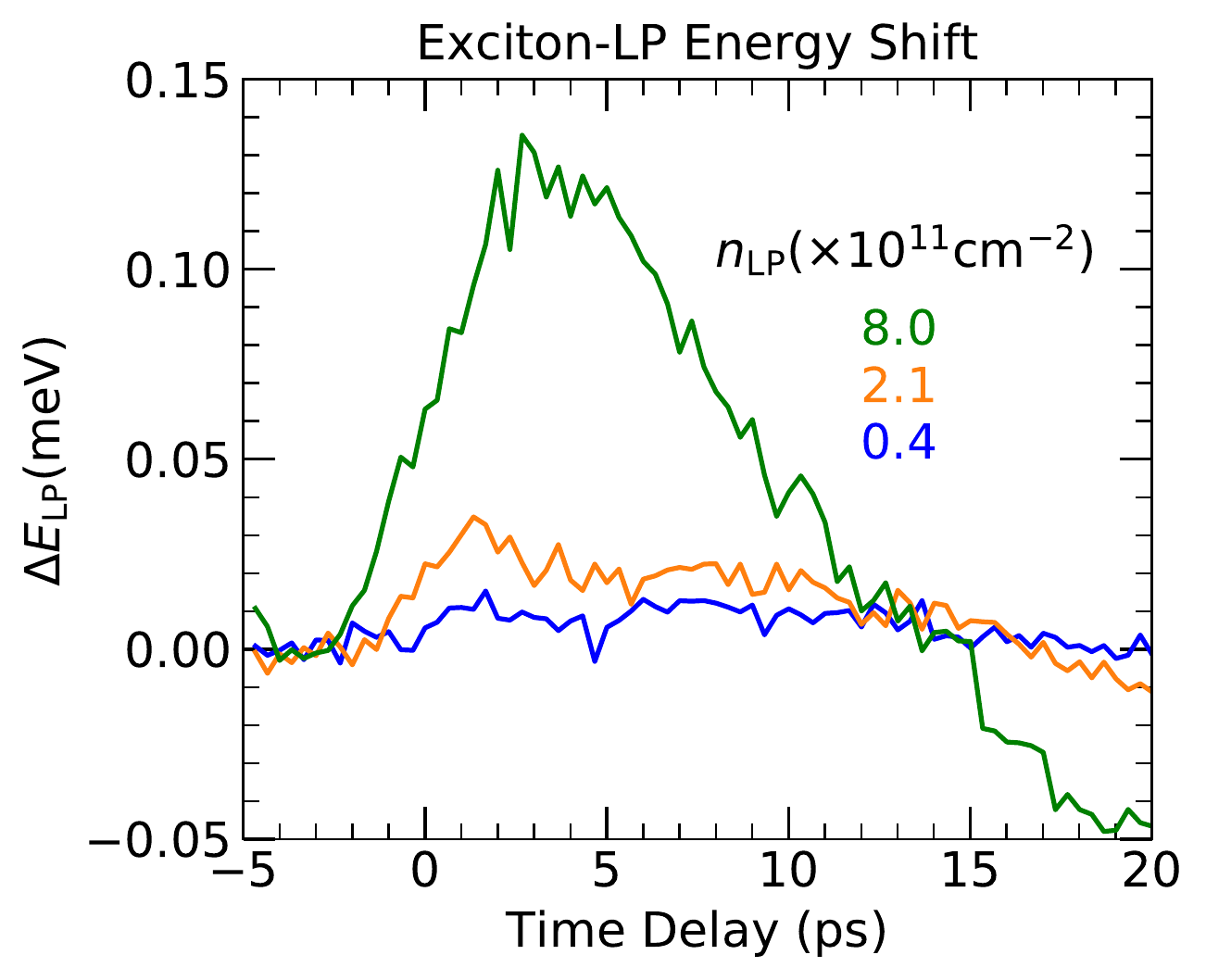}
    \caption{Evolution of the LP energy shift as a function of time delay for different polariton densities.}
\label{fig:excBlueshift}
\end{figure}

We extract $\Delta E_\mathrm{LP}$, the magnitude of the exciton-LP resonance energy shift relative to its $\tau < 0$ value as a function of $\tau$ in Figure~\ref{fig:excBlueshift}. We do not observe clear evidence of effects occurring after the decay of coherent polaritons.

\section{Long-time Scale Pump-Probe}
\begin{figure}
            \includegraphics[width=0.7\textwidth]{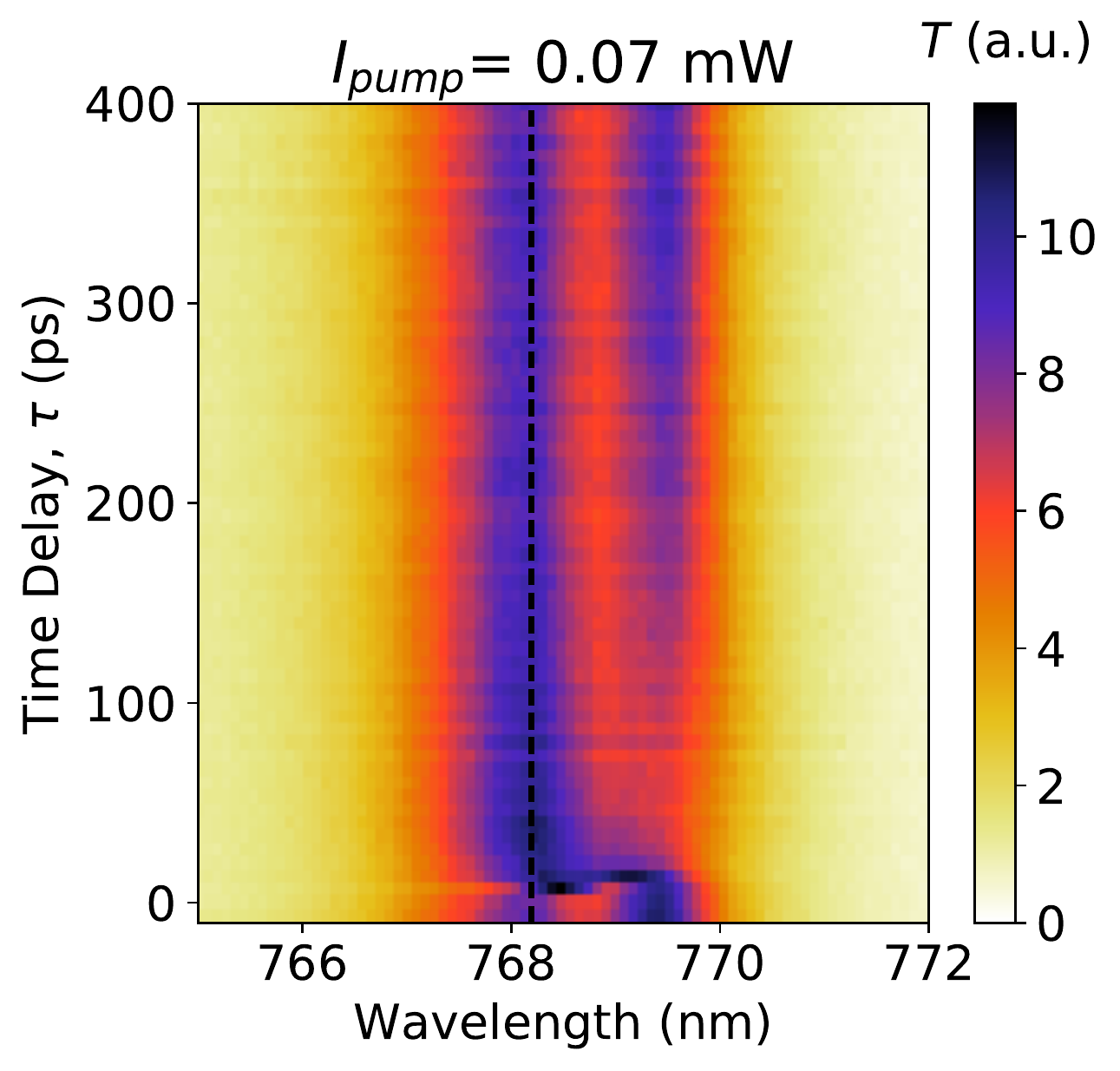}
    \caption{Transmission (T) spectrum of the attractive polaron-polariton when resonantly pumping the UP (indicated by black dashed line) as a function of time delay, $\tau$ from -10 ps to 400 ps at $n_e = (8 \pm 1) \times 10^{11}$ cm$^{-2}$ and $I_{\mathrm{pump}} = 0.07$mW.}
\label{fig:longtau}
\end{figure}
In the pump-probe measurements where we resonantly pump the attractive polaron- LP and the attractive polaron- UP resonantly in Figure~\ref{fig: figure2} and~\ref{fig: figure3}, we consistently observed a shift of the oscillator strength from the attractive polaron- LP to the UP resonance that lasted long after the gain in the transmission was over. We conducted follow-up measurements for long time delay scans (up to $400$ ps) while pumping the attractive polaron- UP. We find the timescale for the recovery of the initial conditions to be $\sim$300 ps (see Fig.~\ref{fig:longtau}).

\end{document}